\documentclass[journal]{IEEEtran}
\usepackage{verbatim}
\usepackage{lineno,hyperref}
\modulolinenumbers[5]
\usepackage{multirow}
\usepackage{ragged2e}
\usepackage{amsmath}
\usepackage{commath}

\usepackage{cite}  
\usepackage{pgfplots}
\usepackage{amsfonts} 
\pgfplotsset{compat=1.3}
\usepackage{soul,xcolor}
\usepackage{tikz}
\usetikzlibrary{arrows,decorations,backgrounds,patterns,matrix,shapes,fit,calc,shadows,plotmarks,pgfplots.groupplots}
\usepackage{graphicx}

\begin{document}
\title{Performance of closed orbit feedback systems with spatial model mismatch}


\author{Sajjad~Hussain~Mirza, Rahul~Singh, Peter~Forck and Bernd~Lorentz
\thanks{S.~H.~Mirza, R.~Singh, P.~Forck and B.~Lorentz are associated with GSI Helmholtzzentrum f\"ur Schwerionenforschung GmbH, Darmstadt, Germany}
\thanks{S.~H.~Mirza, is associated also with Technische Universit\"at Darmstadt, Darmstadt, Germany}}


\maketitle
\begin{abstract}
Closed orbit feedback (COFB) systems used for the global orbit correction rely on the pseudo-inversion of the orbit response matrix (ORM). A mismatch between the model ORM used in the controller and the actual machine ORM can affect the performance of the feedback system. In this paper, the typical sources of such model mismatch such as acceleration ramp ORM variation, intensity-dependent tune shift and beta beating are considered in simulation studies. Their effect on the performance and the stability margins are investigated for both the slow and fast regimes of a COFB system operation. The spectral radius stability condition is utilized instead of the small gain theorem to arrive at the theoretical limits of COFB stability and comparisons with simulations for SIS18 of GSI and experiments at the Cooler synchrotron (COSY) in the Forschungzentrum J\"ulich (FZJ) are also presented.

\end{abstract}
\IEEEpeerreviewmaketitle
\section{Introduction}\label{intro}
\IEEEPARstart{C}losed orbit feedback (COFB) systems are implemented in modern synchrotrons and storage rings to ensure the transverse beam stability against external perturbations as well as local dipolar magnet field errors~\cite{gen,cofbreview,sandira1,nsls,esrf,petra}. Orbit correction relies on the orbit response matrix (ORM) which is defined as the response of the closed orbit to the individual kicks of $M$ corrector magnets (actuators) measured at $N$ beam position monitors (sensors). The closed orbit feedback system design and operation is, therefore, a Multiple-Input-Multiple-Output (MIMO) control problem. The elements of the real-valued ORM of dimension $M \times N$ are given as~\cite{Sands, svd1} 
\begin{equation}\label{eq1}
    R_{mn}= \frac{\sqrt{\beta_{m}\beta_{n}}}{2\sin(\pi Q)}{\cos \left(Q \pi-|\mu_{m}-\mu_{n}|\right)}-{\frac{D_{m}D_{n}}{(\frac{1}{\gamma^2}-\frac{1}{\gamma_t^{2}})C}}
\end{equation} 
where $\beta$, $\mu$ and $D$ denote the beta function, phase advance and dispersion at BPM and corrector locations marked as $m$ and $n$  respectively, in the given plane. $Q$ is the integrated phase advance around the machine (synchrotron or storage ring) referred as the coherent betatron tune of the machine.
$C$ is the circumference of the machine while $\gamma$ is the Lorentz factor.\\ 
For a fixed lattice synchrotron, the parameters of Eq.~\ref{eq1} i.e. $\beta,~\mu$ and $D$ depend solely on the spatial location of the BPMs and correctors, hence the ORM is referred to as the spatial model (fixed over time) of the system in the context of COFB system. 
Temporal response of BPM electronics, corrector magnets and their power supplies etc.  can be represented in continuous frequency domain as a separate transfer function $g(s)e^{-\tau_{d}s}$ (where $s$ represents the Laplace domain variable and $\tau_d$ is the pure delay in seconds) resulting in the total process model as $\mathbf{G}(s)= g(s)e^{-\tau_{d}s}\mathbf{R}$~\cite{sandira1}. Here, it is assumed that all corrector magnets and BPMs have same temporal responses, respectively, which is a realistic assumption in most of the cases.
\begin{figure}[h]
\begin{tikzpicture}
\node[circle,draw, thick, fill=white,inner sep=0pt,minimum size=5pt] (circ1) at (-3.9,1) {$-$};
\node[circle,draw, thick, fill=white,inner sep=0pt,minimum size=5pt] (circ2) at (2.1,1) {$+$};

\node[thick,draw,text width=1.2cm,minimum height=0.8cm,align=center] (controller) at (-2.1,1) {$k(z) \mathbf{R_{\Theta}^{+}}$};
\node[thick,text width=1.2cm,minimum height=0.8cm,align=center] (controller_text) at (-2.1,1.8) {\footnotesize{controller}};
\node[thick,text width=1.2cm,minimum height=0.8cm,align=center] (z0_text) at (-4.6,1.3) {\footnotesize{$w_r=0$}};
\node[thick,text width=1.2cm,minimum height=0.8cm,align=center] (delta_z_text) at (-3.3,1.3) {\footnotesize{$\Delta w(z)$}};
\node[thick,text width=1.2cm,minimum height=0.8cm,align=center] (theta_text) at (-0.85,1.3) {\footnotesize{$\Theta(z)$}};
\node[thick,text width=1.2cm,minimum height=0.8cm,align=center] (z_text) at (1.5,1.3) {\footnotesize{$w(z)$}};
\node[thick,text width=1.2cm,minimum height=0.8cm,align=center] (z_text) at (2.6,1.3) {\footnotesize{$r(z)$}};
\node[thick,text width=1.2cm,minimum height=0.8cm,align=center] (system_text) at (0.4,1.8) {\footnotesize{system}};
\node[thick,text width=1.2cm,minimum height=0.8cm,align=center] (disturbance) at (2.1,2) {\footnotesize{$d(z)$}};
\node[thick,draw,text width=1.2cm,minimum height=0.8cm,align=center] (system) at (0.4,1) {$g(z) \mathbf{R}$};
\draw[-latex',thick] (-5,1)--++(circ1);
\draw[-latex',thick] (controller)--++(system);
\draw[-latex',thick] (system)--++(circ2);
\draw[-latex',thick] (circ1)--++(controller);
\draw[-latex',thick] (2.1,1.8)--++(circ2);
\draw[-latex',thick] (circ2)--++(0.8,0);
\draw[thick] (circ2)--++(0,-1.0);
\draw[thick] (2.1,0)--++(-6,0);
\draw[-latex',thick] (-3.9,0)--++(circ1);
\end{tikzpicture}
\caption{Schematic of the digital closed orbit feedback system in the generic form. $w_r$ is the reference orbit, $\Delta w(z)$ is the error signal, $\Theta (z)$ is the corrector settings, $r(z)$ is the residual orbit and $d(z)$ is the disturbance.}
\label{fig1}
\end{figure}
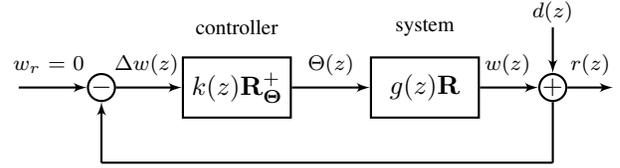
The control system is implemented in ``sample and hold'' mode where the beam position is measured at discrete time steps \{$t=kT_s, k \in \bf{ \mathbb{Z}^+}$\} with $T_s$ being the sampling time interval and the actuator (corrector magnets) inputs are held constant over the time interval $t \in \{kT_s, (k+1)T_s$\}. The corresponding zero-order hold and sampled discrete-time model of $g(s)e^{-\tau_{d}s}$ can be represented with the help of modified $z$-transform ($\mathcal{Z}_m$) as~\cite{sandira1, ducane, sami}
\begin{equation}\label{eq1_1}
    g(z)= z^{-l}(z-1) \mathcal{Z}_m \left[\frac{g(s)}{s}\right]
\end{equation}
where $l$ is the smallest positive integer satisfying $lT_{s} \geq \tau_d $ while $g(s)$ is a first-order low pass filter of bandwidth $a$ (rad/s) and is given as 
\begin{equation}\label{eq1_2}
    g(s)= \frac{a}{s+a}
\end{equation}
Consequently, the controller is also modelled as having two separate parts; temporal part $k(z)$ as well as a spatial part such that the total model $\mathbf{K(z)}= k(z)\mathbf{R_{\Theta}^{+}}$, 
where $\mathbf{R_{\Theta}^{+}}$ is the pseudo-inverse of $\mathbf{R}_{\Theta}$. The closed-loop block diagram of such a feedback system is shown in Fig.~\ref{fig1}. The subscript $\Theta$ is used to distinguish the controller model ORM from the actual system ORM $\mathbf{R}$, as their relative discord is the main focus of this paper. The required corrector settings $\Theta(z)$ for a given perturbed orbit vector $\Delta w(z)$ are calculated as,
\begin{equation}\label{eq2}
    \Theta (z)= k(z)\mathbf{R_{\Theta}^{+}}\Delta w (z)
\end{equation}
where $\Delta w$ represents the transverse orbit displacement relative to design orbit in either plane (i.e. $w$ is either $x$ or $y$). 
Singular value decomposition (SVD)~\cite{svdgolub} is the most widely used method for the pseudo-inversion of a real valued ORM by its decomposition in one diagonal matrix $\mathbf{S}_{\Theta}$ and two orthogonal matrices $\mathbf{U}_{\Theta}$ and $\mathbf{V}_{\Theta}$ as $\textbf{R}_{\Theta}=\textbf{U}_{\Theta} \textbf{S}_{\Theta} \textbf{V}_{\Theta}^{\text{T}}$ (here the superscript $\text{T}$ represents the transpose). 
The elements of the diagonal matrix $\mathbf{S}_{\Theta}$ called singular values, $\sigma_i$ are the characteristic of the matrix $\mathbf{R}_{\Theta}$ and their inverses act as the controller gain for each mode (column) of the orthogonal matrices $\textbf{U}_{\Theta},\textbf{V}_{\Theta}$.
For the symmetric and the near-symmetric arrangements of BPMs and corrector magnets, a DFT-based diagonalization and inversion method is recently discussed in the context of closed orbit correction~\cite{self2} as an alternate to SVD resulting in a sparse representation for changing lattices. These special matrices will be utilized in this report for deriving exact relations due to their favourable properties.\\
A mismatch between the actual ORM of the synchrotron $\mathbf{R}$ and its model used in the controller $\mathbf{R}_{\Theta}$ can have a detrimental effect on the performance and the stability of the COFB system. The possible reasons for such a model mismatch fall into two broad categories: a) Uncertainty in the measurement of the orbit response matrix due to BPM and corrector calibration errors~\cite{sandira2}  or b) change in the orbit response matrix due to drift of the machine settings during operation~\cite{self3}, tune shift being the most popular one.
Typical sources of tune shift include magnet current drifts during long storage of the beams, magnet gradient errors during the fast ramps~\cite{self3, bonn} and intensity-dependent tune shifts at higher beam intensities~\cite{RS2}. Our focus in this contribution is on the latter type since it cannot be improved by calibration and is generally less discussed in the literature. Further, ramping hadron synchrotrons such as SIS18 and SIS100 of the Facility of Antiproton and Ion Research (FAIR) have a variation of the ORM on the acceleration ramp by design along with an increase in beam rigidity during the acceleration cycle. These conditions pose additional challenges for the implementation of the COFB system and require a thorough study of the COFB robustness against the large spatial model mismatch.
It is however important to note that, if the time scale of the ORM variation due to aforementioned effects is longer (an order of magnitude) than the delays and bandwidths of the temporal response of the BPMs and corrector magnets, the ORM can still be considered as a separate part $\mathbf{R}$ in the total system model $\mathbf{G}(s)$. Therefore the ORM variations are treated exclusively as the spatial model mismatch in this contribution.\\
The previous work reported on this topic has primarily focused on stability conditions from the small gain theorem (SGT)~\cite{sgt,cernuncern} which yields a conservative stability criterion and then on the enhancement of the stability margins for higher-order modes of the ORM using the Tikhonov regularization~\cite{sandira1}. In this paper, we show that the conservative stability margin yielded by SGT certainly ensures the closed orbit stability but is too tight in comparison to the observed limits in simulations and experiments. Instead of SGT, we have obtained the stability limits using spectral radius stability condition~\cite{multivariable}. The motivation is to establish the theoretical stability margins consistent with the simulation and experimental results against the various sources of model mismatch. To our knowledge, this is the first experimental demonstration of the closed orbit stability limits due to the spatial model mismatch. 
Further the sensitivity of the feedback system to the lattice and control parameter variations is studied, in order to answer design questions like a) how many ORMs need to be updated in the controller during the on-ramp orbit correction when the machine ORM changes over time, b) variation of which lattice parameters is crucial e.g. tune (global lattice parameter) or beta function at the location of BPMs and correctors (local lattice parameter) (See Eq.~\ref{eq1}).\\
The orbit correction systems are divided into two temporal regimes, based upon the relative time scale of the correction rate (reciprocal of time step for update of controller output) and the dynamics of the system. For the slow feedback systems, the time step of orbit correction is at least an order of magnitude higher than than system delays and open-loop time constant (the time required to achieve $63.2\%$ of the steady-state). Systems, where the correction rate is comparable to temporal dynamics are referred to here as the fast feedback systems. The performance of a COFB system against the spatial model mismatch is primarily evaluated by the correction ratio in the slow regime and system bandwidth in the fast regime. Correction ratio is the ratio of residual orbit to original orbit after one correction while bandwidth is characterized by the frequency range of effective orbit correction i.e. frequency of disturbances attenuation up to $-3$dB.

In the next sections,  the spectral radius is presented along with a discussion on the relation between the instability thresholds predicted by the spectral radius and the SGT for the special case of circulant ORMs for the slow feedback regime. The measurements of spectral radius corresponding to the inherent model mismatch are also discussed. Following that, the spectral radius is calculated for typical model mismatch scenarios in GSI SIS18~\cite{accelerators}. The conclusions drawn from the calculations and simulations for SIS18 are verified experimentally at COSY FZJ~\cite{cosy}. The analysis is extended to the fast feedback regime where it is shown that the temporal bandwidth and the controller gains of the COFB system can be expressed as the functions of the spectral radius. Spectral radius is thus put forward as a general figure of merit for the optimal operation of the COFB systems in the presence of a model mismatch. 
\section{Spatial model mismatch in slow temporal regime}\label{definematrix}
\begin{figure}[ht]
\begin{tikzpicture}
\node[circle,draw, thick, fill=white,inner sep=0pt,minimum size=5pt] (circ1) at (-3.9,1) {$-$};
\node[circle,draw, thick, fill=white,inner sep=0pt,minimum size=5pt] (circ2) at (2.1,1) {$+$};

\node[thick,draw,text width=1.2cm,minimum height=0.8cm,align=center] (controller) at (-2.1,1) {$\mathbf{R_{\Theta}^{+}}$};
\node[thick,text width=1.2cm,minimum height=0.8cm,align=center] (controller_text) at (-2.1,1.8) {\footnotesize{controller}};
\node[thick,text width=1.2cm,minimum height=0.8cm,align=center] (delta_z_text) at (-3.3,1.3) {\footnotesize{$-r(z)$}};
\node[thick,text width=1.2cm,minimum height=0.8cm,align=center] (theta_text) at (-0.85,1.3) {\footnotesize{$\Theta(z)$}};
\node[thick,text width=1.2cm,minimum height=0.8cm,align=center] (z_text) at (1.5,1.3) {\footnotesize{$w(z)$}};
\node[thick,text width=1.2cm,minimum height=0.8cm,align=center] (z_text) at (2.6,1.3) {\footnotesize{$r(z)$}};
\node[thick,text width=1.2cm,minimum height=0.8cm,align=center] (system_text) at (0.4,1.8) {\footnotesize{system}};
\node[thick,text width=1.2cm,minimum height=0.8cm,align=center] (disturbance) at (2.1,2) {\footnotesize{$d(z)$}};
\node[thick,draw,text width=1.2cm,minimum height=0.8cm,align=center] (system) at (0.4,1) {$z^{-1} \mathbf{R}$};
\draw[-latex',thick] (controller)--++(system);
\draw[-latex',thick] (system)--++(circ2);
\draw[-latex',thick] (circ1)--++(controller);
\draw[-latex',thick] (2.1,1.8)--++(circ2);
\draw[-latex',thick] (circ2)--++(0.8,0);
\draw[thick] (circ2)--++(0,-1.0);
\draw[thick] (2.1,0)--++(-6,0);
\draw[thick] (-3.9,0)--++(circ1);
\end{tikzpicture}
\caption{Schematic of the digital closed orbit feedback system in the slow regime. $-r(z)$ is the error signal, $\Theta (z)$ is the corrector settings, $r(z)$ is the residual orbit and $d(z)$ is the disturbance defined in Eqs.~\ref{eq2_3} to~\ref{eq2_5}.}
\label{fig1_1}
\end{figure}
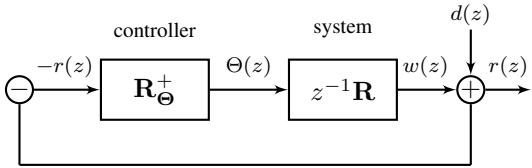

The slow temporal regime of orbit correction is a scenario where the temporal dynamics of the system (corrector magnets and their power supplies etc.) is fast enough to be ignored as compared to the rate of update of the controller output. An example is the correction of a static orbit distortion caused by a fixed non-oscillating magnet field error in a synchrotron. In this case, the sampling time $T_s$ can be selected in such a way that the effect of each controller output is fully settled and the system is in the steady-state before the next corrector input. The orbit correction for low frequency perturbations with fast corrector magnets (having bandwidth an order of magnitude higher than frequency of interest) can also be categorized as slow regime of orbit correction. \\
The closed orbit feedback system in the slow regime is implemented in a way that the effect of corrector settings $\Theta_k$ calculated at a given time step \{$t=kT_s, k \in \bf{ \mathbb{Z}^+}$\} corresponding to the perturbed orbit of previous time step $r_{k-1}$ is added to $r_{k-1}$ i.e.,
\begin{equation}\label{eq2_1}
\centering
    r_{k} = r_{k-1} + \mathbf{R}\Theta_{k}
\end{equation}
where $\Theta_{k}=-\mathbf{R_{\Theta}^{+}}r_{k-1}$. Consequently $r_k$ can be written as 
\begin{equation}\label{eq2_2}
\centering
    r_{k} = r_{k-1} - \mathbf{R}\mathbf{R_{\Theta}^{+}}r_{k-1} = \left(\textbf{I}-\textbf{R}{\textbf{R}^{+}_\Theta}\right)r_{k-1}
\end{equation}
Here, $\bf{I}$ is the identity matrix of the size of $\bf{RR_{\Theta}^{+}}$ 
which is a square matrix even if $\mathbf{R}$ and $\textbf{R}^{+}_\Theta$ are rectangular matrices, since $r_k$ and $r_{k-1}$ are vectors of same dimension equal to the number of BPMs.
Converting the difference Eq.~\ref{eq2_2} into $z$-domain, one obtains  
\begin{equation}\label{eq2_3}
\centering
    r(z) = z^{-1}r(z)-z^{-1}\mathbf{R}\mathbf{R_{\Theta}^{+}}r(z)
\end{equation}
Since $g(s)=1$ and $\tau_d << T_s$ for slow regime, following Eq.~\ref{eq1_1} $g(z)=z^{-1}$ while $k(z)=1$ and for a constant initial perturbation, the residual from previous time step acts as the disturbance to be removed at the current time step such that i.e. $d(z) = z^{-1}r(z)$. The corresponding block-diagram is shown in Fig.~\ref{fig1_1}  for which Eq.~\ref{eq2_3} can be re-written as

Since $g(s)=1$ and $\tau_d << T_s$ (for slow regime), following Eq.~\ref{eq1_1} $g(z)=z^{-1}$ and $k(z)=1$ and for a constant initial perturbation, the residual from previous time step acts as the disturbance to be removed at the current time step such that i.e. $d(z) = z^{-1}r(z)$. The corresponding block-diagram is shown in Fig.~\ref{fig1_1}  for which Eq.~\ref{eq2_3} can be re-written as
\begin{eqnarray}\label{eq2_5}
\centering
    r(z) &=& d(z)-z^{-1}\mathbf{R}\mathbf{R_{\Theta}^{+}}r(z) \nonumber \\
         &=&  [\mathbf{I}+z^{-1}\mathbf{R}\mathbf{R_{\Theta}^{+}}]^{-1}  d(z) \nonumber \\
         &=&  [\mathbf{I}+\mathbf{L}]^{-1}  d(z) \nonumber\\
         &=&  \mathbf{\zeta}(z)d(z)
\end{eqnarray}
The term $\mathbf{\zeta}(z)$ is by definition the sensitivity function from disturbance to the output for the closed loop of a MIMO system where $\mathbf{L}$ is the loop transfer function equal to the product of system and controller models. \\
From Eq.~\ref{eq2_2}, the residual orbit after $k$ time steps with respect to initial perturbed orbit $r_{-1} = w_0$ can be written as,  
\begin{equation}\label{eq4}
    \centering
    r_\text{k} = \left(\textbf{I}-\textbf{R}{\textbf{R}^{+}_\Theta}\right)^k w_0
\end{equation}
It is evident from Eq.~\ref{eq4} that for the slow regime, the spatial model mismatch can be fully described by the following combined matrix structure referred to as ``correction matrix", 
\begin{equation}\label{eq5}
    \centering
    \textbf{M} = \textbf{I}-\textbf{R}{\textbf{R}^{+}_\Theta}
\end{equation}
which will be a null matrix for a perfect model match i.e. for ${\textbf{R}_\Theta} = \textbf{R}$.
\subsection{Stability criteria for the slow regime of COFB system}
Equations~\ref{eq4} and~\ref{eq5} show that the matrix $\textbf{M}$ has to be a convergent matrix for the residual orbit to approach zero after subsequent time steps. The condition of convergence is given as 
\begin{equation}\label{eq7}
    \centering
     \rho(\textbf{M}) < 1
\end{equation}
where $\rho(\bf{M})$ is known as the spectral radius of matrix $\bf{M}$ and is defined as the largest absolute eigenvalue of $\textbf{M}$ (A.$5.3$ of~\cite{multivariable}) as
\begin{equation}\label{eqnonumber1}
    \centering
    \rho(\textbf{M}) = \max_i\{|\lambda_i|\} \nonumber
\end{equation} 
Inequality~\ref{eq7} can be regarded as the general stability criteria for a COFB system in the slow regime against any source of spatial model mismatch as it depends only on the relative discord between $\textbf{R}$ and $\textbf{R}_\Theta$ and can be helpful to find the limits of optical parameter (e.g. tune) variations beyond which the correction will lead to the orbit divergence and the beam loss.\\
It is important to mention that pseudo-inverse $\textbf{R}^{+}_\Theta$ in Eq.~\ref{eq4} is often truncated for numerical stability and for robust closed orbit correction, which leads to a reduction in the rank of $\textbf{R}^{+}_\Theta$. In this case, the matrix $\mathbf{R}$ also needs to be equally truncated before the calculation of $\rho(\textbf{M})$ which otherwise will be misleadingly equal to $1$ because of the truncated modes of $\textbf{R}^{+}_\Theta$. 
\subsection{Comparison to small gain theorem}
\begin{figure}[ht]
\centering
\begin{tikzpicture}
\node[thick,draw,text width=1cm,minimum height=1cm,align=center] (delta) at (0,2.7) {$\Delta$};
\node[thick,draw,text width=1cm,minimum height=1cm,align=center] (T) at (0,1) {\textbf{T}};
\draw[thick] (delta)--++(-1.5,0);
\draw[latex-,thick] (delta)--++(1.5,0);
\draw[thick] (-1.5,2.7)--++(0,-1.7);
\draw[thick] (1.5,2.7)--++(0,-1.7);
\draw[latex-,thick] (T)--++(-1.5,0);
\draw[thick] (T)--++(1.5,0);
\end{tikzpicture}
\caption{System interconnection for the application of small gain theorem. \label{fig1a}}
\end{figure}
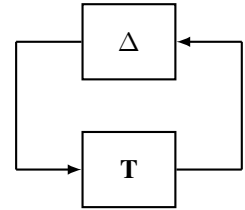

Typically, the small gain theorem (SGT) is used to find the stability margins of the COFB system~\cite{cernuncern,sandira2} which is a sufficient condition for the closed-loop stability but yields very stringent margins. SGT relies on expressing the uncertainty in the MIMO process model as a separate matrix in the structure shown in Fig.~\ref{fig1a} where $\Delta$ is the uncertainty and the $\textbf{T}$ is the rest of the MIMO transfer function seen by the $\Delta$. The condition of stability given by SGT theorem for the interconnection of Fig.~\ref{fig1a} is given as $||\Delta||_{\infty} \cdot ||\textbf{T}||_{\infty} < 1$ where $||\cdot||_{\infty}$ is the $\textbf{H}_{\infty}$ norm\footnote{defined as the highest singular value of the matrix.} (Sec.~$4.10.2$ of~\cite{multivariable}).
Here we make a theoretical comparison between SGT and spectral radius condition discussed in inequality~\ref{eq7} based upon following assumptions: 1) we consider  circulant or block-circulant ORMs (a scenario which exists for FAIR synchrotrons and also some synchrotron light sources~\cite{self2}) for which the spectral radius is equal to the $\textbf{H}_{\infty}$ norm i.e. the highest singular value $\bar{\sigma}$. 2) we consider only the well conditioned ORMs for which the general inverse exists such that $\textbf{R}_\Theta{\textbf{R}^{+}_\Theta}=\mathbf{I}$. 3) we assume only the additive uncertainty in the ORM in order to deal with ORM errors arising from optical parameter variations e.g. tune and beta function. It is important to mention that these assumptions are required only for a clear comparison between two methods. The spectral radius condition itself holds also for non-quadratic and non-symmetric ORMs as the matrix $\bf{M}$ is always a square matrix.  


For errors of additive nature in ORM, the SGT based stability criterion has been worked out in~\cite{cernuncern} and ignoring temporal part (i.e. taking $\textbf{T} = \mathbf{R}^{+}_\Theta)$, it is written as
\begin{eqnarray}
\label{eq9}
    \centering
     \bar{\sigma}(\Delta_R)&<&\frac{1}{\bar{\sigma}(\mathbf{R}^{+}_\Theta)}\\ \nonumber
     \bar{\sigma}(\Delta_R)&<&\underline{\sigma}(\mathbf{R}_\Theta)
\end{eqnarray}
where $\Delta_R = \mathbf{R}-\mathbf{R}_{\Theta}$ and $\underline{\sigma}$ is the smallest singular value.
The spectral radius stability condition shown in inequality~\ref{eq7} can also be solved in a similar manner for circulant symmetric ORMs as,
\begin{equation}
\label{eqnonumber2}
    \centering
    \rho( \textbf{I}-\textbf{R}{\textbf{R}^{+}_\Theta})=\rho(-\Delta_R\mathbf{R}^{+}_\Theta) = \rho(\Delta_R\mathbf{R}^{+}_\Theta)=\bar{\sigma}(\Delta_R\mathbf{R}^{+}_\Theta) < 1 \nonumber
\end{equation}
Using the singular value inequality $\bar{\sigma}(A)\underline{\sigma}(B)\leq \bar{\sigma}(AB)$ (A.$3.4$ of~\cite{multivariable}), 
\begin{eqnarray}
\label{eq10}
    \centering
\bar{\sigma}(\Delta_R) \underline{\sigma}(\mathbf{R}^{+}_\Theta) < \bar{\sigma}(\Delta_R\mathbf{R}^{+}_\Theta) &<&1\\ \nonumber
    \bar{\sigma}(\Delta_R)&<&\frac{1}{\underline{\sigma}(\mathbf{R}^{+}_\Theta)}\\ \nonumber
    \bar{\sigma}(\Delta_R)&<&\bar{\sigma}(\mathbf{R}_\Theta)
\end{eqnarray}
The highest allowed uncertainty is limited by the smallest singular value of the nominal ORM in case of SGT while the spectral radius condition yields a much higher margin i.e. of the highest singular value. The ratio between the two margins is exactly the condition number of the nominal ORM for the circulant symmetry case.   
\subsection{Measurement of the spectral radius}
The $\textbf{H}_\infty$ norm of matrix $\textbf{M}$ of Eq.~\ref{eq5} can be defined as (Sec.~$3.3.2$ of~\cite{multivariable})
\begin{equation}
\label{eq11}
    \centering
   ||\textbf{M}||_{\infty} =max\frac{||r_1||_2}{||w_0||_2} \geq \frac{r_{1,\text{RMS}}}{w_{0,\text{RMS}}} = \delta_1
\end{equation}
where $||\cdot||_2$ is the Euclidean norm\footnote{defined as $||x||_2 = \sqrt{x_{1}^2 + x_{2}^2 + ....x_{n}^2 }$}. For the circulant symmetry, $\rho(\textbf{M}) = ||\textbf{M}||_{\infty}$ and the measured correction ratio can be related with $\rho(\textbf{M})$ as
\begin{equation}
\label{eq12}
    \centering
      \rho(\textbf{M}) \geq \delta_1 
\end{equation}
For non-circulant ORMs, $\rho(\textbf{M})\leq||\textbf{M}||_{\infty}$ (A.$5.3$ of~\cite{multivariable})
and the correction ratio is asymptotically related to $\rho(\textbf{M})$ by Gelfand's formula~\cite{lax} $$\rho(\textbf{M})=\lim_{k\to\infty} ||\textbf{M}^k||_{\infty}^{\frac{1}{k}}$$ as
\begin{equation}
\label{eq13}
    \centering
    \rho(\textbf{M}) \geq \lim_{k\to\infty} \frac{r_{\text{k,RMS}}}{r_{\text{k-1,RMS}}} = \lim_{k\to\infty} \delta_k
\end{equation}
In general, an approximation $\rho(\textbf{M}) = \frac{r_{1,\text{RMS}}}{w_{0,\text{RMS}}}$ can be made if the largest effect of model mismatch is on the dominant mode in correction space.  As discussed in the next sections, this is a reasonable approximation in practical situations.
\section{Application to the SIS18 COFB system: simulations}\label{sis18}
\begin{figure}[ht]
    \centering
    \includegraphics*[width=8.35cm]{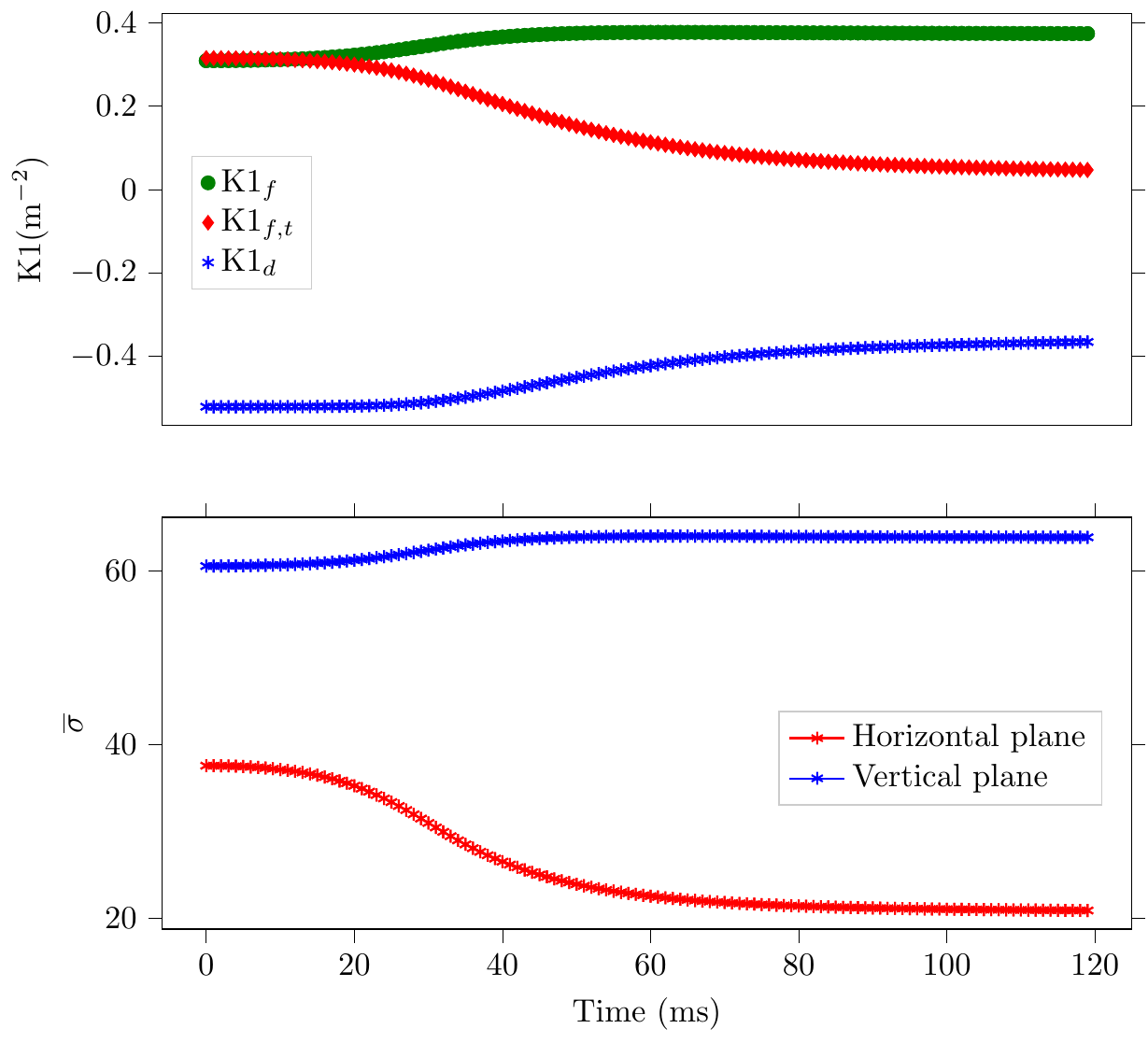}
    \caption{Top: Triplet to doublet quadrupole transition over the ramp of 10 T/s in SIS18 (an example of optics change during acceleration cycle). $K1_f, K1_d$ and $K1_{f,t}$ are the normalized strengths of the doublet focusing, doublet defocusing and triplet focusing quadrupole families of SIS18, respectively. Bottom: The variation of the highest singular values of the ORM in both transver planes, over the ramp of 10 T/s.}
    \label{fig2}
\end{figure}

The SIS18 synchrotron is used to accelerate a wide variety of ions (from protons to Uranium) to a variable range of energies and will serve as the booster ring to the SIS100 synchrotron of the FAIR facility~\cite{FAIR,Spiller}. A COFB system is under development at SIS18 which should operate from injection till extraction including the acceleration ramp~\cite{self1} for the beam quality preservation before injection into SIS100. Triplet focusing configuration is used at beam injection in order to keep the transverse machine acceptance large enough to accommodate multi-turn injection~\cite{tdb}. During the acceleration, triplet quadrupole strength is gradually minimized leaving only the doublet focusing by the end of acceleration ramp. The resulting quadrupole settings variation during a typical dipole magnets ramp (10 T/s) of SIS18 is shown in Fig.~\ref{fig2} (top) where the normalized quadrupole strengths ($K1 = \frac{1}{B\rho} \frac{\partial B}{\partial z}$ in $m^{-2}$) of focusing, defocusing and triplet focusing quadrupole families are plotted.
The triplet to doublet transition, results into a continuous ORM variation with time~\cite{self2} as evident from the variation of the highest singular values of the ORMs during the ramp plotted for both planes in Fig.~\ref{fig2} (bottom), although the design (set) tune remains the same. The singular values are calculated in MAD-X~\cite{madx} using the time-dependent quadrupole settings of the ramp. Further, tune shifts at injection in high current settings as well as on-ramp tune shifts~\cite{RS1} caused by magnet gradient errors~\cite{RS2} have been observed experimentally. In this section, we calculate the spectral radius and simulate the correction ratio in MAD-X as a function of model mismatch of various sources to estimate instability thresholds of the SIS18 COFB system, in the slow feedback regime. The ideas are later extended to the fast feedback regime in section~\ref{temporal}.    
\subsection{Spectral radius and correction ratio due to on-ramp model variation}
\begin{figure}[ht]
    \centering
    \includegraphics*[width=8.3cm]{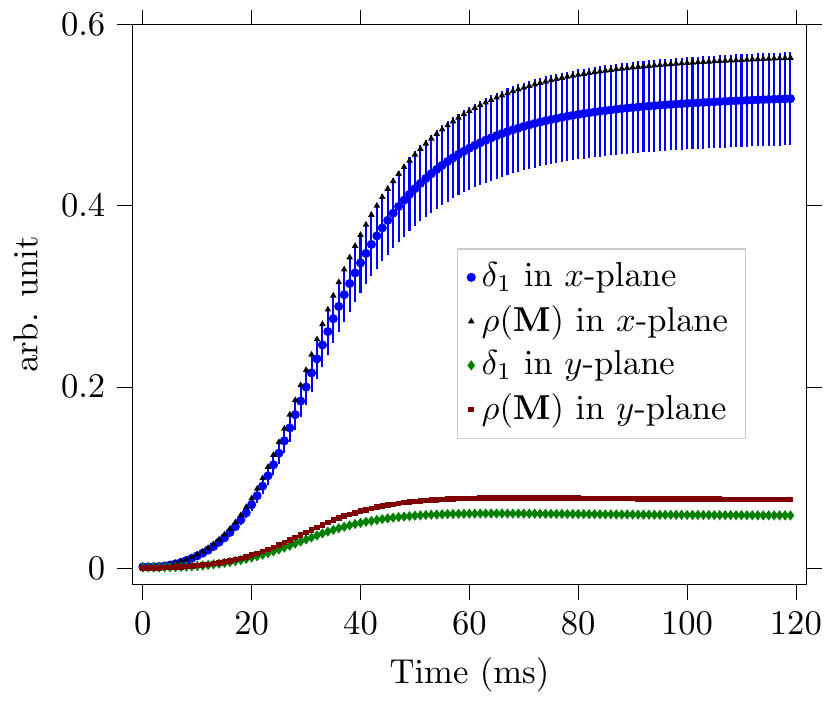}
    \caption{The simulated correction ratio $\delta_1=\frac{r_{1_{\text{RMS}}}}{\Delta z_{0_{\text{RMS}}}}$ and $\rho(\textbf{M}(t))$ over a ramp of 10 T/s for both planes.}
    \label{fig6}
\end{figure}

\begin{figure}[ht]
    \centering
    \includegraphics*[width=8.3cm]{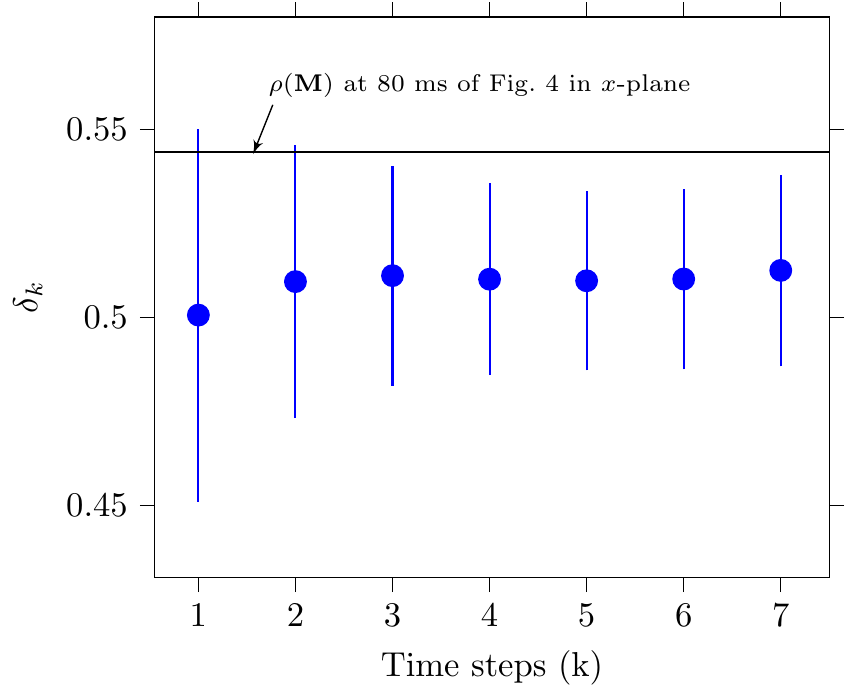}
    \caption{Asymptotic behavior of the standard deviation of correction ratio $\delta_k = \frac{r_{k,RMS}}{r_{k-1,RMS}}$ over subsequent iterations in agreement with Gelfand's formula.}
    \label{fig7}
\end{figure}

Figure~\ref{fig2} (bottom) shows that the ORM variation in $x$-plane is more prominent ($\approx$ 45$\%$ of initial value) than that in y-plane ($\approx$ 6$\%$) for SIS18.
In order to check the possibility of using a fixed-ORM based COFB system over the ramp, the spectral radius $\rho(\mathbf{M})$ defined between the ORMs $\mathbf{R(t)}$ at each time step of the ramp (10 T/s) and the nominal ORM $\mathbf{R}_\Theta$ corresponding to the injection settings is given as

    \begin{equation}\label{eq14}
    \centering
    \rho(\textbf{M}(t)) =  \rho(\textbf{I}-\textbf{R}(t){\textbf{R}^{+}_\Theta})
\end{equation}
The result of Eq.~\ref{eq14} is plotted in Fig.~\ref{fig6} for both planes (red for $x$-plane and black for $y$-plane). The simulated correction ratio $\delta_1=\frac{r_{1,\text{RMS}}}{\Delta z_{0,\text{RMS}}}$ is also plotted in Fig.~\ref{fig6} as a function of time (blue for $x$-plane and green for $y$-plane). The dots in the plots comprises of the mean RMS of the Gaussian distribution of 1000 corrected orbits at each time step while the error bar represents the standard deviation. The initial perturbed orbits were
obtained by random misalignment of all $24$ focusing and defocusing quadrupoles of SIS18. The details of the simulation strategy 
are discussed in detail in~\cite{self3}. In the comparison of correction ratio to the $\rho(\textbf{M}(t))$, one can see that the mean value of the first iteration residual increases in a direct proportionality to the $\rho(\textbf{M})$ (according to inequality~\ref{eq12}) with the increase of model mismatch in both planes confirming that $\rho(\textbf{M})$ is a practical figure of merit for quantifying the spatial model mismatch. Anyhow, a few orbits in $x$-plane have correction ratio larger than $\rho(\textbf{M})$  (at the edge of standard deviation) which is a consequence of slight deviation from circulant symmetry in this plane while a pure circulant symmetry exists in the $y$-plane. The correction ratio is further simulated  in $x$-plane over successive iterations of orbit correction as $\delta_k$ for a fixed model mismatch at $80 $~ms of the ramp where $\rho(\textbf{M}) = 0.544$. The resultant $\delta_k$ is plotted in Fig.~\ref{fig7} for the first $7$ iterations after which the residual reaches to the noise level. One can see that the correction ratio of all simulated orbits is less than $\rho(\textbf{M}) $ asymptotically as predicted by inequality~\ref{eq13} for non-circulant ORMs.   
Moreover, the simulations suggest that the SIS18 COFB should never be unstable in the slow regime, because of the on-ramp spatial model mismatch for both planes. The only difference is that more ORMs will be needed to be updated in the $x$-plane in comparison to the $y$-plane, to achieve a comparable orbit correction in both planes.  
 
\subsection{Intensity dependent tune shift}\label{sis18tune}
\begin{figure}[ht]
    \centering
    \includegraphics*[width=8.3cm]{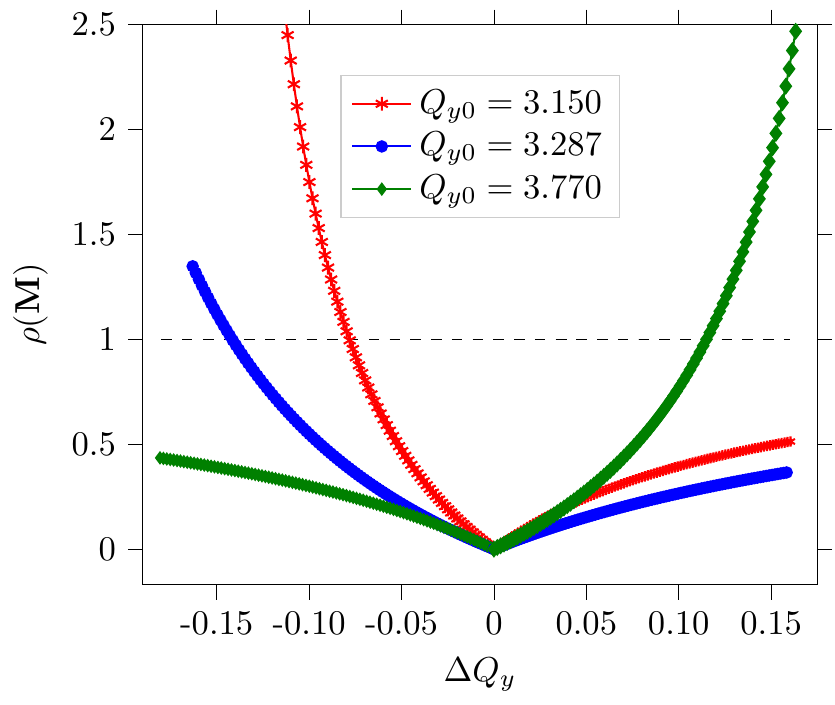}
    \caption{$\rho(\textbf{M}(\Delta Q_y))$ for image charge based tune shift for three different initial vertical tunes.}
    \label{fig9}
\end{figure}

\begin{figure}[ht]
    \centering
    \includegraphics*[width=8.3cm]{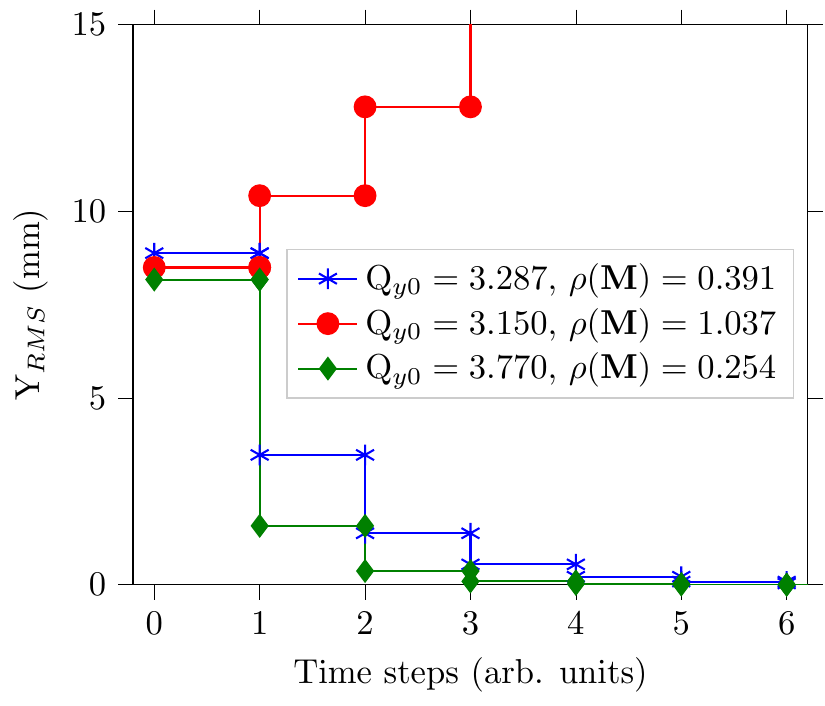}
    \caption{Simulated closed orbit RMS over successive time steps of correction in MAD-X for a fixed $\Delta Q=-0.079$ in the slow regime of SIS18.}
    \label{fig9a}
\end{figure}

The intensity-dependent tune shift can be modelled as the image charge effect of the vacuum chambers around the beam~\cite{ZOTTER}. It is simulated in MAD-X by replacing all the drift regions in SIS18 with weak defocusing quadrupoles in $y$-plane and the same quadrupole strength is added to the strengths of already present quadrupole families resulting in a weak defocusing force throughout the synchrotron.
As a result, we calculate the $\rho(\textbf{M}(\Delta Q))$ similar to that defined in Eq.~\ref{eq14} where $\textbf{R}(t)$ is replaced by $\textbf{R}(\Delta Q)$ corresponding to image charge tune shifts for the vertical plane of SIS18.
 $\overline{\sigma}(\Delta_{R})$ (discussed in Eq.~\ref{eq10}) quantifies the relative deviation of the singular values corresponding to the dominant (tune) modes of $\mathbf{R_{\Theta}}$ and $\mathbf{R}$. This implies that the allowed uncertainty in the ORM for a stable COFB will depend upon the initial value of the tune and the direction of the tune shift towards or away from the integer value. This is due to the fact that the singular values have different growth rate when the tune moves towards (faster) or away (slower) from the integer value.
In order to confirm this, three working points $Q_{y0}= 3.287,~3.150$ and $3.770$ are selected to calculate the ORMs corresponding to the zero tune shifts. The tune of the machine is then changed in both directions for all three cases and $\rho(\textbf{M})$ is plotted in Fig.~\ref{fig9}.
For an initial working point of 3.15 (red), the instability threshold is reached with a negative tune shift of only -0.0782 while for the $Q_{y0} = 3.77$ even a negative tune shift of -0.180 makes $\rho(\textbf{M})=0.43$ (far from instability threshold of $1$). On the other hand, $\rho(\textbf{M})$ crosses the instability threshold for a small positive tune shift for $Q_{y0} = 3.77$. \\
Furthermore, a fixed tune shift of $-0.079$ is selected for all three above mentioned initial tunes and the orbit correction over many time steps is simulated in MAD-X for the slow regime of SIS18. The corrector settings for each time step $k$ are calculated as $\Theta_k = \Theta_{k-1}-\textbf{R}_\Theta^{+} y_k$ where $\textbf{R}_\Theta^{+}$ is the pseudo-inverse of ORM corresponding to initial tune while the actual machine tune in MAD-X is $Q_{y0}-0.079$, for each case.
The RMS values of the resultant closed orbits over each time step are plotted in Fig.~\ref{fig9a} where $k=0$ corresponds to the initial perturbed orbits.
One can see that the orbit correction is unstable for the case of initial tune $Q_{y0}=3.15$ in agreement with $\rho(\textbf{M}) > 1$ for $\Delta Q=-0.079$, as evident in Fig.~\ref{fig9}. For the other two cases, where the $\rho(\textbf{M})$ is less than $1$, the closed orbit necessarily converges to zero with the required number of time steps being proportional to the magnitude of $\rho(\textbf{M})$.
These calculations confirm that it is not only the magnitude of the tune shift that defines the instability threshold of the COFB system but also the direction towards or away from the integer value. Tune shifts away from the integer value ensure the COFB system stability. It is important to mention that COFB system instabilities due to spatial model mismatch are discussed for scenerios when the machine is itself stable against tune resonances.
\subsection{Beta beating} 
Beta beating is another source of model error that can vary the ORM of the machine.
It is the deviation of the beta function from its periodic values caused by the localized quadrupole field errors in a synchrotron and is defined for random field errors $\Delta K_i$ as~\cite{bb},
\begin{eqnarray}
\label{eq15}
    \Delta \beta(l_0) &=&  \frac{\beta_0}{2\sin(2\pi Q)} \sum_{i}^{} \int_{0}^{L} \Delta K_i \beta_i(l) \\ \nonumber &\cdot& \cos\left(2\left(\mu(l)- \mu_0\right)-2 \pi Q\right) dl
\end{eqnarray}
while the corresponding tune variation is  
\begin{equation}\label{eq16}
    \Delta Q = \frac{1}{4\pi} \sum_{i}^{} \int_{0}^{L} \Delta K_i \beta(l) dl
\end{equation}
Here $L$ is the length of the quadrupole and $l$ is the longitudinal coordinate of the synchrotron. The other parameters are already defined in sec.~\ref{intro}.
\begin{figure}[ht]
    \centering
    \includegraphics*[width=8.3cm]{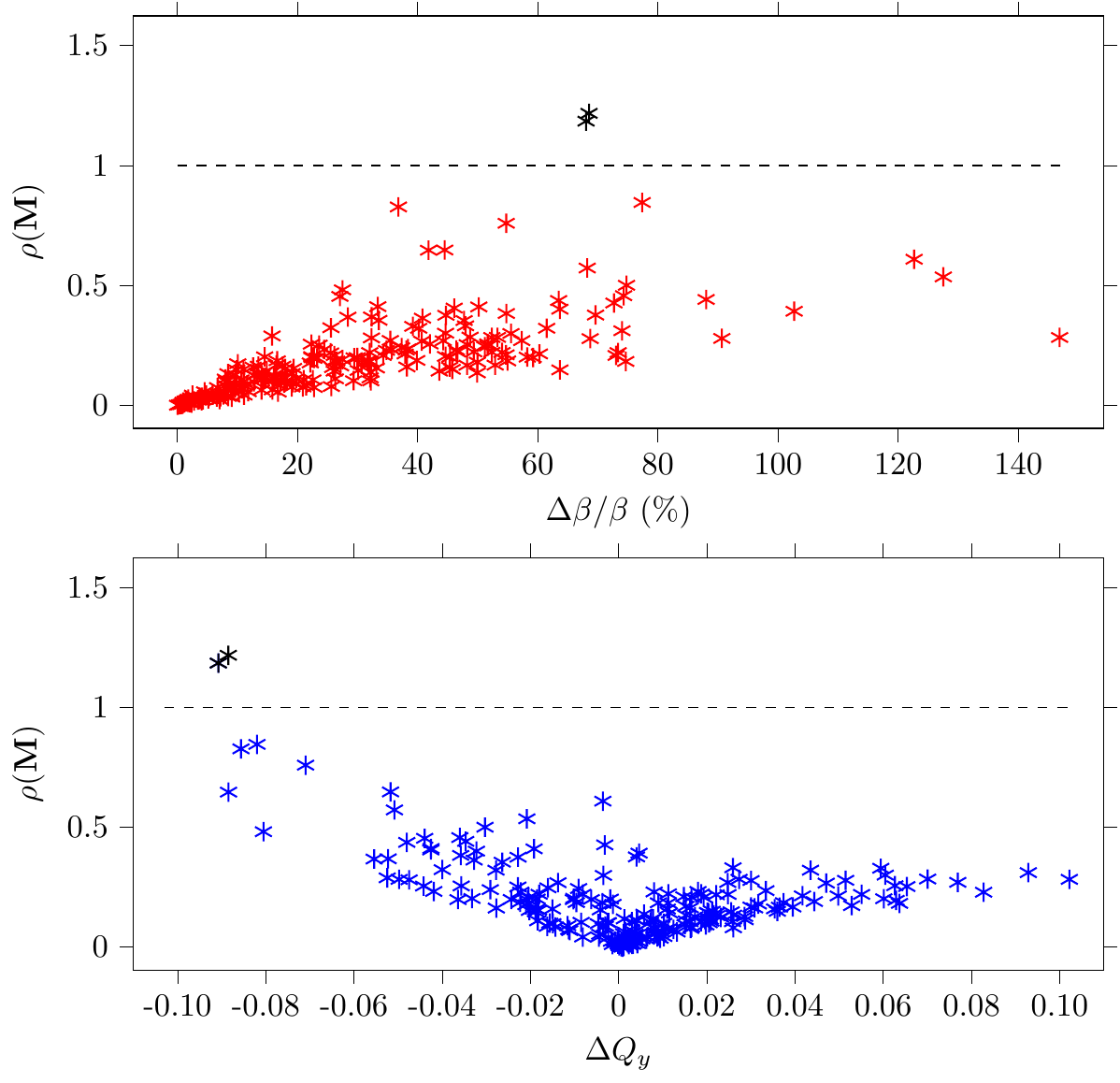}
    \caption{$\rho(\textbf{M}(\frac{\Delta \beta}{\beta}))$ for beta beating produced by randomly distributed field errors for an initial vertical tune $Q_{y0} = 3.287$. Top: Versus beta beating. Bottom: Versus tune shift.}
    \label{fig8_1}
\end{figure}
\begin{figure}[ht]
    \centering
    \includegraphics*[width=8.3cm]{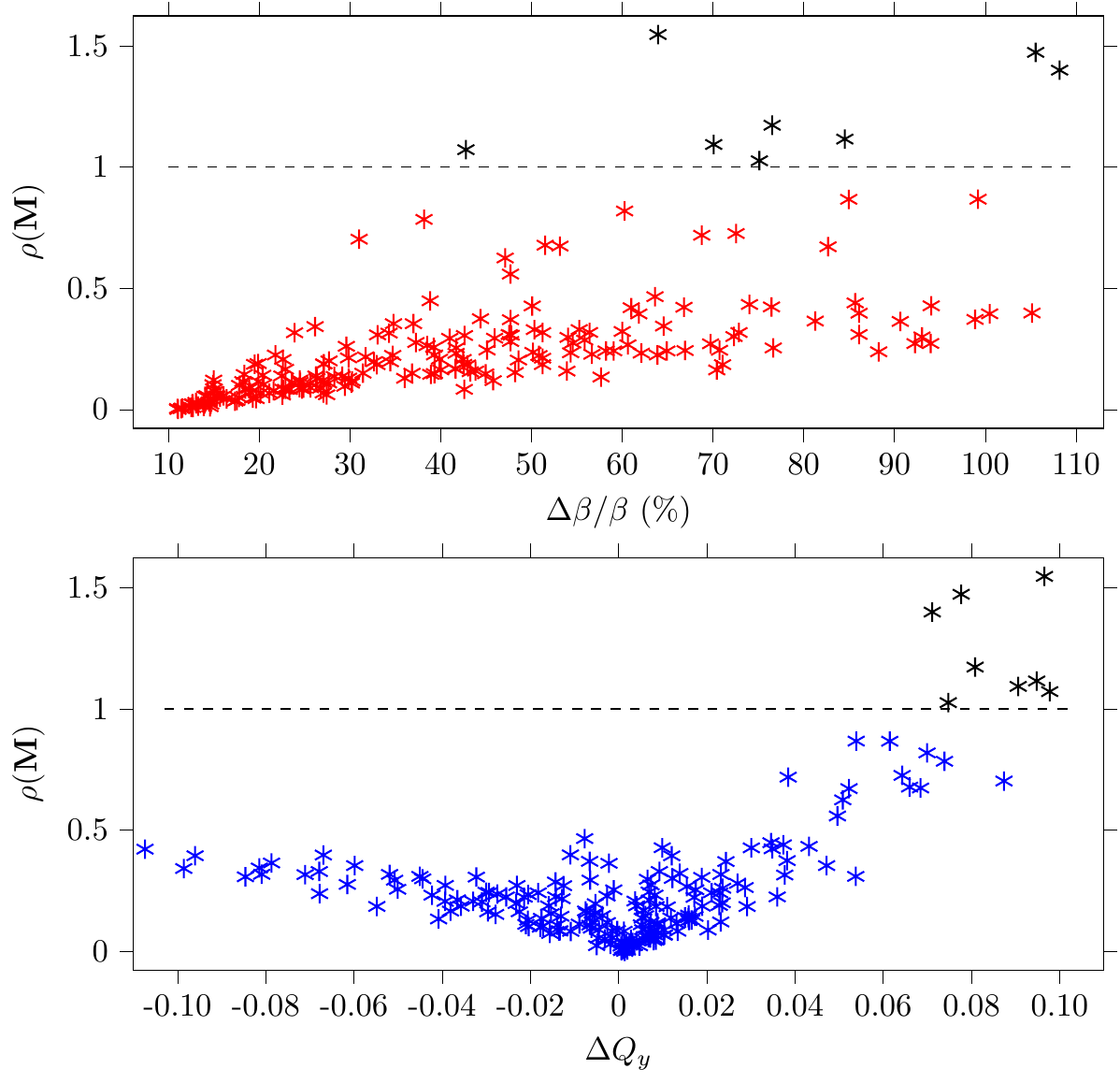}
    \caption{$\rho(\mathbf{M}(\frac{\Delta \beta}{\beta}))$ for beta beating produced by randomly distributed field errors for an initial vertical tune $Q_{y0} = 3.770$. Top: Versus beta beating. Bottom: Versus tune shift.}
    \label{fig9_1}
\end{figure}
Figure~\ref{fig8_1} (top) shows the simulations for $y$-plane of $\rho(\textbf{M}(\frac{\Delta \beta}{\beta}))$ using $\mathbf{R_{\Theta}}$ corresponding to zero beta beating and $Q_{y0} =  3.287$ for a range of beta beating up to $125\%$ generated by random distribution of quadrupole field errors in MAD-X. One can see that the orbit correction with a wrong model, can lead to instability ($\rho(\textbf{M}) >$ 1) even for 67\% of peak-peak beta beating (black stars in Fig.~\ref{fig8_1}) and on the other hand, can converge ($\rho(\textbf{M}) < $ 1) even for more than 100\% of peak-peak beta beating. This observation is explained in Fig.~\ref{fig8_1} (bottom) where $\rho(\textbf{M})$ is plotted as a function of tune shift. The $\rho(\textbf{M})$ is larger than $1$ only for negative tune shifts.
The reason is that the negative tune shift takes the working point of the machine (in MAD-X) closer to the integer value as also discussed in the previous subsection.
This assertion is confirmed by adjusting the unperturbed vertical tune at $ Q_{y0}=3.77$ so that for positive tune shifts, the working point moves towards the integer value (contrary to previous case).  Figure~\ref{fig9_1} shows the corresponding $\rho(\textbf{M})$ as function of both beta beating and tune shift. The behavior is similar to the previous case but mirrored with respect to tune shift. Thus we conclude, that the magnitude of beta beating itself does not play a large role in context of orbit correction but only when accompanied by the tune shift towards the integer value, beta beating can lead to orbit instability.
\section{Experiments at COSY FZJ}\label{cosy}
Some of the conclusions drawn in sec.~\ref{sis18} were experimentally verified at the COFB system of proton Cooler Synchrotron at Forschungzentrum J\"ulich (FZJ). The correction time for the operational COFB system is $T_s= 2 s$ and therefore, it can be regarded as a slow feedback system with temporal characteristics of the power supplies and BPMs ignored.
The on-ramp ORM variation and artificially introduced tune shift were used as the sources of model mismatch. By default, the singular values of the ORM in the controller having magnitude below the $10\%$ of the highest singular value were truncated.
\subsection{Vertical tune variation}
\begin{figure}[ht]
    \centering
    \includegraphics*[width=8.3cm]{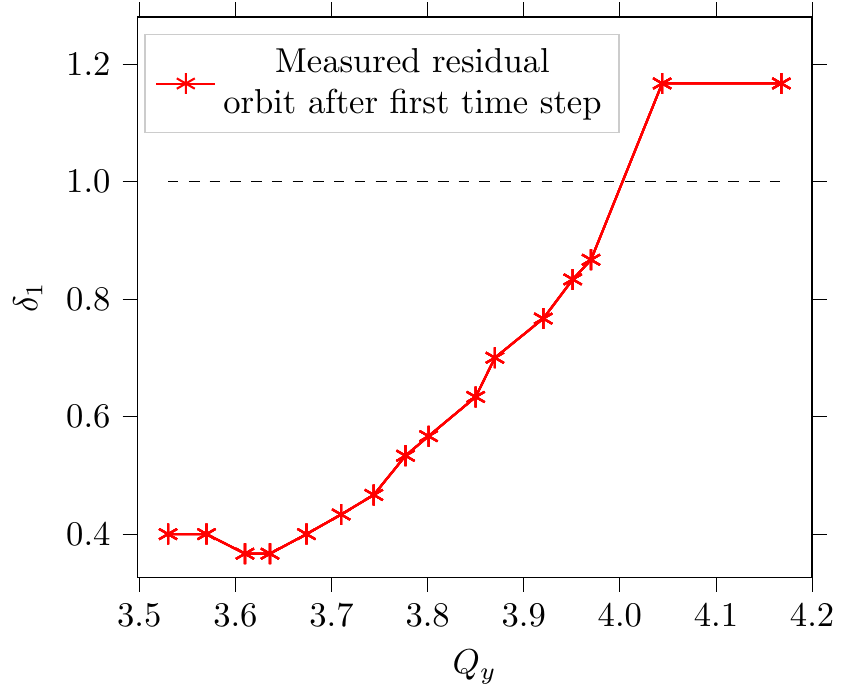}
    \includegraphics*[width=8.3cm]{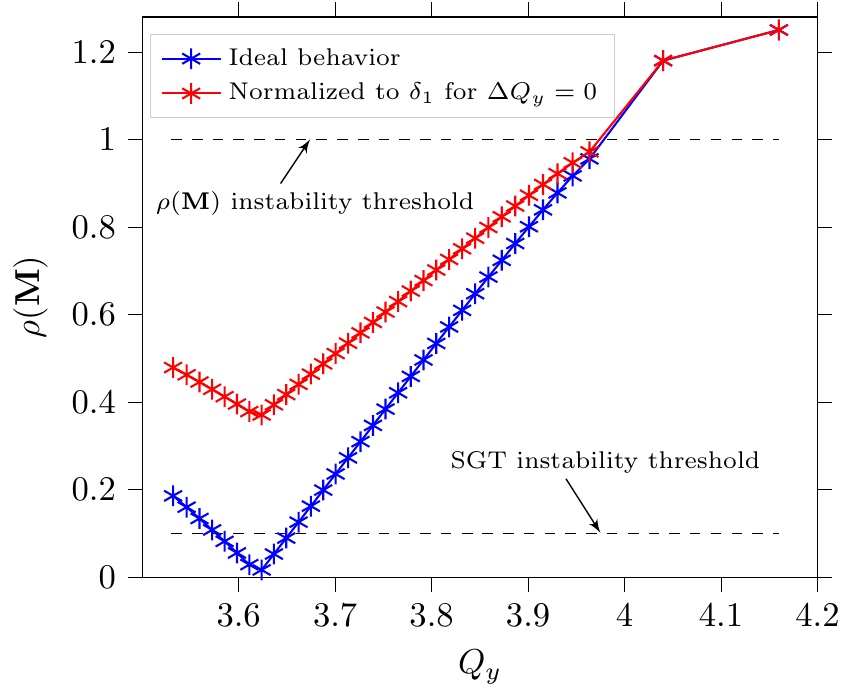}
    \caption{\label{fig15}Measured ratio $\delta_1=\frac{r_{1_{\text{RMS}}}}{\Delta y_{0_{\text{RMS}}}}$ in the vertical plane of COSY for the fixed machine tune of $Q_{y0}=3.62$ while varying the ORM in the feedback system corresponding to a tune range of $3.53$ to $4.16$ (top). Calculated $\rho(\mathbf{M})$ for vertical plane (bottom). For ideal and realistic case please see the text.}
\end{figure}

As concluded in sec.~\ref{sis18tune}, it is not only the magnitude of the tune shift but its direction towards or away from integer tune that defines the instability threshold of the COFB system. In order to verify this experimentally, the machine settings of COSY corresponding to the vertical tune of $Q_{y0} =3.62$ were fixed while the ORMs were updated in the feedback system for a range of tune values from $3.53$ to $4.16$. The resultant measured correction ratio ($\delta_1=\frac{r_{1,\text{RMS}}}{\Delta z_{0,\text{RMS}}}$) is plotted in Fig.~\ref{fig15} (top). One can see that the correction ratio increases as the tune moves towards the integer value. Figure~\ref{fig15}~(bottom) shows the corresponding calculated $\rho(\textbf{M})$ in blue which is largely different from measured correction ratio primarily around $\Delta Q_y=0$ showing an inherent model mismatch between the actual (unknown) model $\textbf{R}_{X}$ and the best known (measured) model $\textbf{R}$ of the machine. Since $\textbf{R}$ is the known model in this case, we can condition it before pseudo-inversion by SVD that $\textbf{R}\textbf{R}^{+}=\bf{I}$. Such a model mismatch can be associated to the corrector calibration errors resulting in a multiplicative uncertainty such that $\mathbf{R}_{X} = (\mathbf{I}+\Delta_\mathbf{R})\mathbf{R}$. 
The effect of such an inherent model mismatch can be excluded from simulations using the measured correction ratio for $\Delta Q_y=0$. Let us call it $r'_{1}$ 
\begin{eqnarray}
\label{eq17}
    \centering
    r'_{1} &=& \left(\textbf{I}-\textbf{R}_{X}{\textbf{R}^{+}}\right) w_0  \\ \nonumber
     r'_{1} &=& -\Delta_\mathbf{R} w_0 
\end{eqnarray}
The effect of intentionally varied model $\textbf{R}_{\Theta}$ (controller model) on the residual $r_1$ with respect to the actual ORM $\textbf{R}_{X}$ can be written as
\begin{eqnarray}\label{eq18}
    \centering
    r_\text{1} &=& \left(\textbf{I}-\textbf{R}_{X}\textbf{R}^{+}_{\Theta}\right) w_0  \nonumber \\
    r_\text{1} &=& \left(\textbf{M}-\Delta_\mathbf{R}\left(1-\textbf{M}\right)\right) w_0
\end{eqnarray}
where $\mathbf{M}=(\mathbf{I}-\textbf{R}\textbf{R}^{+}_{\Theta})$. Using inequality~\ref{eq12} and the triangular inequalities of matrix norm ($||\mathbf{A}+\mathbf{B}|| \leq ||\mathbf{A}||+||\mathbf{B}||$ and $||\mathbf{A}\mathbf{B}|| \leq ||\mathbf{A}||\cdot||\mathbf{B}||$) the correction ratio can be deduced from Eq.~\ref{eq18} as 
\begin{equation}
\label{eq19}
    \centering
    \frac{r_\text{1,RMS}}{w_\text{0,RMS}} \leq \rho(\textbf{M})+\frac{r'_\text{1,RMS}}{w_\text{0,RMS}}\left(1-\rho(\textbf{M})\right)
\end{equation}
The right-hand side of inequality Equation~\ref{eq19} is plotted in red in Fig.~\ref{fig15}~(bottom) in order to match the experimentally measured correction ratio. 
\begin{figure}[ht]
    \centering
    \includegraphics*[width=8.3cm]{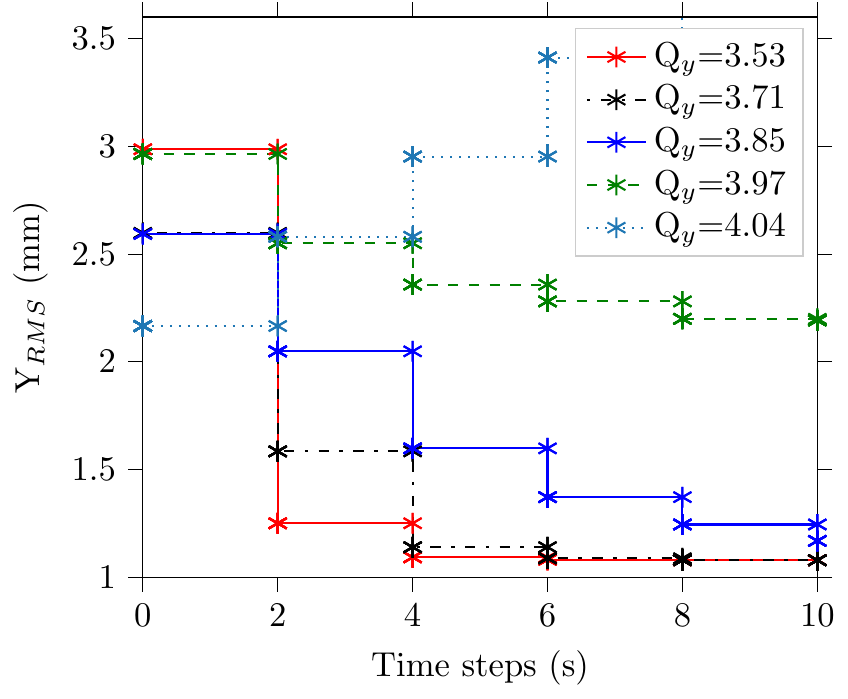}
    \caption{\label{fig16}Measured RMS of the closed orbit in $y-$plane of COSY synchrotron for different ORMs used for the corrector settings calculation while keeping the machine settings fixed for a tune of 3.62.}
\end{figure}
Figure~\ref{fig16} shows the closed orbit RMS over a number of time steps for specific tune values. Even up to a very large tune shift ($Q_{y} = 3.97$), the COFB system is stable since $\rho(\textbf{M})$ is still less than $1$ (Fig.~\ref{fig15} bottom). For the tune shifts $Q_y > 4$, the $\rho(\textbf{M})>1$ and the COFB system gets unstable as seen by the increasing residual over successive time steps in Fig.~\ref{fig16} resulting in the beam loss.
  
\subsection{On-ramp ORM variation}
\begin{figure}[ht]
    \centering
    \includegraphics[width=8.3cm]{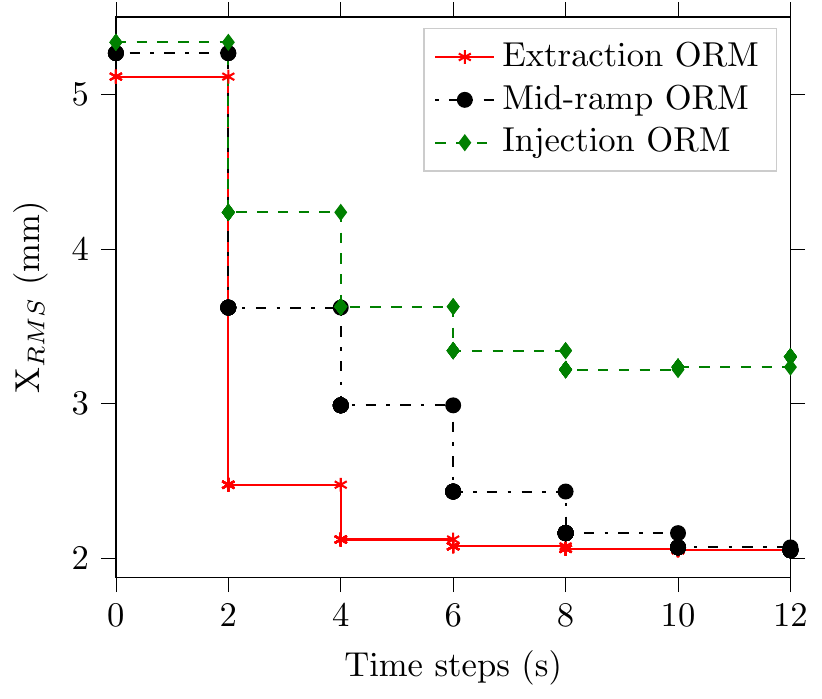}
    \includegraphics[width=8.3cm]{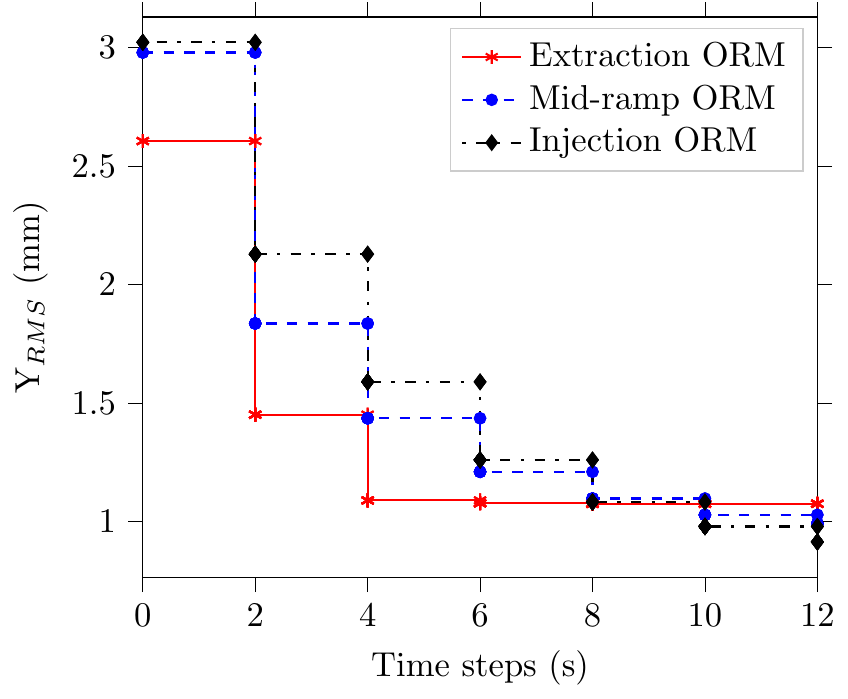}
    \caption{\label{fig10}Measured RMS of the closed orbit in $x$-plane (top) and in $y-$ plane (bottom) for machine settings corresponding to injection, mid-ramp and extraction energy while using the ORM corresponding to extraction settings for the corrector settings calculation.
   }
\end{figure}

Orbit corrections were performed in both planes at injection, mid-ramp and extraction energies using only the ORM corresponding to the extraction settings. The temporal evolution of the closed orbit RMS is plotted as the function of time steps in Fig.~\ref{fig10} (top) for $x$-plane and (bottom) for $y$-plane for all three machine settings. 
One can see that the residual after the first time step as well as the number of correction steps required for the orbit convergence increase with the increased model mismatch. In all three cases, closed orbit converges to a minimum, confirming that the COFB system stability is ensured if the correction ratio is less than 1, in agreement with the spectral radius stability condition of inequality~\ref{eq7}. Two observations can be made concerning the residual orbit. 1) A residual orbit remains after the first correction step even at extraction due to the inherent model mismatch, as discussed in the previous subsection. 2) The residual orbit over successive correction steps converges to different values for the injection, mid ramp and extraction. This is most likely due to the change in orbit pattern and a different coupling to the truncated SVD modes in $\mathbf{R}_{\Theta}$ in each of those cases. 
\section{Temporal response with model mismatch}\label{temporal}
In deriving the stability condition in the slow feedback regime i.e. inequality~\ref{eq7}, we had ignored the temporal parts of the system and the controller. Here, we extend the treatment to the fast regime where the system dynamics plays a role in the performance of the COFB system. The notion of the required number of discrete time steps for orbit convergence in slow regime can be exchanged with the closed-loop bandwidth for the fast regime. This is intuitive in a way that larger the value of $\rho(\textbf{M})$, higher is the residual $r_1$ after first orbit correction and more time steps or longer time constant is required for the convergence of the closed orbit. The longer time constant is equivalent to the reduction of the achievable bandwidth of the COFB system for the attenuation of the dynamic orbit errors.\\
Referred to Fig.~\ref{fig1}, the sensitivity function from disturbance to output is used for describing the bandwidth and the loop transfer function and complementary sensitivity function for the closed loop are used for determining the stability margins of the COFB system, in this section. The standard sensitivity function $\zeta(z)$~\cite{multivariable} for our system can be written by definition as
\begin{eqnarray}
\label{eq20}
    \centering
     \mathbf{\zeta}(z) & \overset{\Delta}{=}& [\mathbf{I}+g(z)\mathbf{R} k(z) \mathbf{R}_{\Theta}^{+}]^{-1}  \nonumber \\
     &\overset{\Delta}{=}& [\mathbf{I}+\mathbf{I}g(z) k(z)-(\mathbf{I}-\mathbf{R}\mathbf{R}_{\Theta}^{+})g(z) k(z)]^{-1} 
\end{eqnarray}
The matrix $\mathbf{M}$ appears in the temporal sensitivity function in order to affect the achievable bandwidth. 
Eq.~\ref{eq20} is a general equation of sensitivity function which can be solved for each mode of the ORM and for any controller $k(z)$. 
Hereon we calculate the sensitivity functions $\zeta_{\rho} (s)$ corresponding to the highest eigenvalue i.e. $\rho(\textbf{M})$ in order to demonstrate the largest effect of model mismatch on the bandwidth, since all other eigenvalues will be smaller than $\rho(\textbf{M})$. In other words, we are approximating the MIMO system with a SISO system for $\rho(\textbf{M})$.
Replacement of $\mathbf{M}$ with $\rho(\mathbf{M})$ and rearrangement of Eq.~\ref{eq20} yields
\begin{equation}
\label{eq21}
    \centering
     \mathbf{\zeta}_\rho(z) = \frac{\frac{1}{1+g(z)k(z)}}{1-\rho(\mathbf{M}) \frac{g(z)k(z)}{1+g(z)k(z)}}
\end{equation}
For the first order system model $\frac{a}{s+a}$ given in Eq.~\ref{eq1_2} with the delay $e^{-\tau_d s}$, $g(z)$ can be calculated using Eq.~\ref{eq1_1} as 
\begin{equation}
\label{eq21_1}
    \centering
     g(z)=z^{-l} \frac{b_{0}z + b_1}{z-a_{1}}
\end{equation}
where $l$ is the delay in the units of sampling time $T_s$, $a_1=e^{-aT_s}$, $b_0=1-e^{a(\tau_{d}-lT_s)}$ and $b_1=e^{a(\tau_{d}-lT_s)}-e^{-aT_s}$. For $k(z)$ (the controller for $\rho(\mathbf{M})=0$) we use the Internal Model Controller (IMC) methodology~\cite{IMC} in which the controller is based upon the invertible part of the system model augmented by a low pass filter $Z(z)= z^{-l} \frac{b_{0}z + b_1}{z-a_{1}}$ and can be calculated by equating the function $\frac{g(z)k(z)}{1+g(z)k(z)}$ of Eq.~\ref{eq21} to the low pass filter $Z(z)$ (following the literature~\cite{michael}) and is given as 
\begin{equation}
\label{eq22}
    \centering
     k(z) = [g(z)]^{-1}\frac{Z(z)}{1-Z(z)}
\end{equation}
IMC-based controller can also be approximated as classical PI controller for the first order systems\cite{IMC_PI_course,IMC_PI}. Using Eqs.~\ref{eq21_1} and~\ref{eq22}, the $\rho(\mathbf{M})$-dependent sensitivity function of Eq.~\ref{eq21} can be written as, 
\begin{equation}\label{eq23}
    \centering
     \mathbf{\zeta}_\rho(z) = \frac{1-Z(z)}{1-\rho(\mathbf{M})Z(z)} 
\end{equation}
The corresponding loop transfer function $L_\rho(z)$ can be calculated from its following relation to the sensitivity function 
\begin{equation}\label{eq24}
    \centering
    L_\rho(z) = \dfrac{1-\zeta_\rho(z)}{\zeta_\rho(z)}
\end{equation}
where the comparison of Eq.~\ref{eq23} and~\ref{eq24} yields
\begin{equation}\label{eq25}
    \centering
    \begin{split}
     {L}_\rho(z) =  \frac{Z(z)\left(1-\rho(\mathbf{M})\right)}{1-Z(z)} \\
     = L_{(\rho=0)}(z)\left(1-\rho(\mathbf{M}\right)
    \end{split}
\end{equation}
where $L_{(\rho=0)}(z)$ is the loop transfer function when the spatial models of system and controller cancel each other i.e.
\begin{equation}\label{eq26}
    \centering
     L_{(\rho=0)}(z) = g(z)k(z)
\end{equation}
Comparison of Eqs.~\ref{eq25} and~\ref{eq26} yields the $\rho(\mathbf{M})$-dependence of the controller parameters as, 
\begin{equation}\label{eq27}
    \centering
     {k}_\rho (z) = k(z)\left(1-\rho(\mathbf{M})\right)
\end{equation}
with reference to $k(z)$, the controller corresponding to $\rho(\textbf{M})=0$.
\subsection{Temporal bandwidth}
\begin{figure}[ht]
    \centering
    \includegraphics*[width=8.3cm]{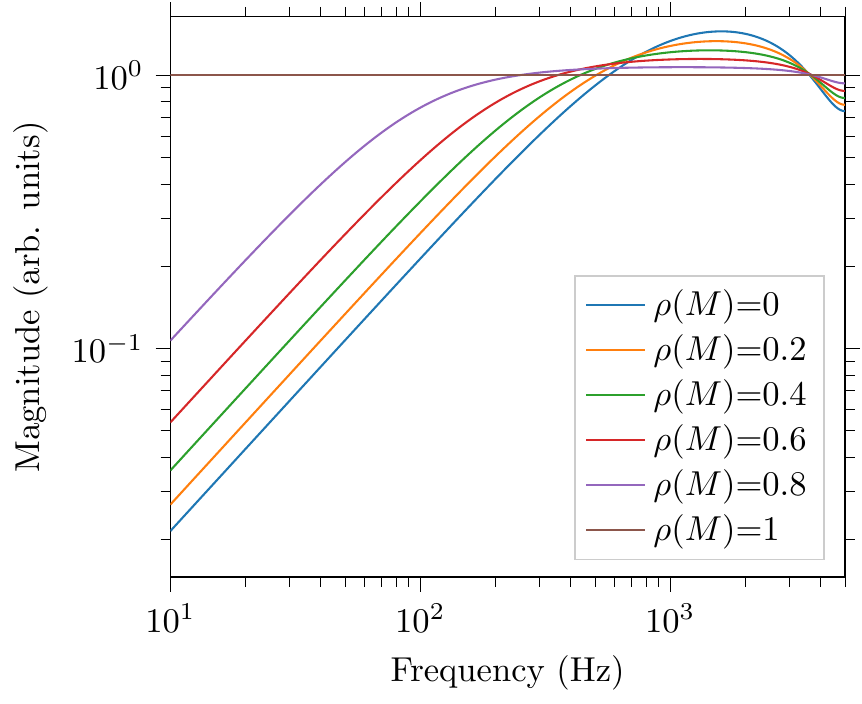}
    \includegraphics*[width=8.3cm]{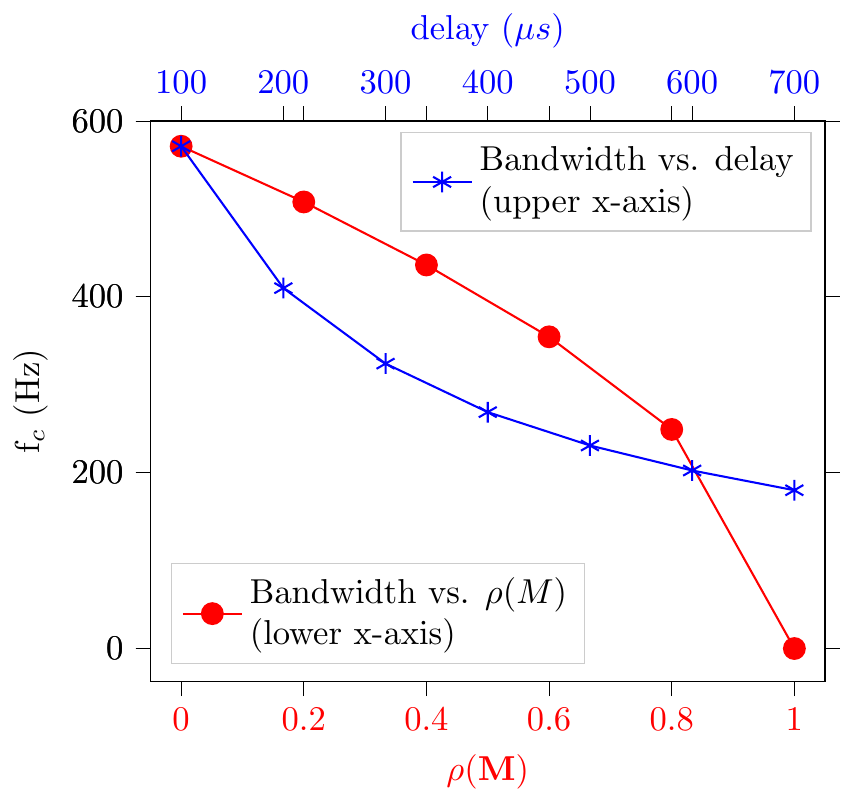}
    \caption{Top: Plot of sensitivity function of internal model controller (IMC) vs. frequency for various values of $\rho(\mathbf{M})$ between $0$ and $1$. Bottom: Reduction in the cut-off frequency as a function of  $\rho(\mathbf{M})$ (lower $x$-axis) and delay (upper $x$-axis)}.
    \label{fig3}
\end{figure}

The sensitivity function $\mathbf{\zeta}_\rho(z)$ defined in Eq.~\ref{eq23} is plotted for a range of $0 < \rho(\textbf{M}) < 1$  for a low pass filter $Z(z) = z^{-1} \frac{0.413}{z-0.586}$ calculated for the bandwidth of $850$~Hz, $\tau_d = 100~\mu s$ and $T_s = 100~\mu s$ in Fig.~\ref{fig3}~(top).
One can see that there is no disturbance rejection for $\rho(\textbf{M}) = 1$. The bandwidth decreases with the increase of $\rho(\textbf{M})$ up to $1$ as plotted in Fig.~\ref{fig3} (bottom in red). For a comparison, the bandwidth vs. temporal delay in a range of $\tau_d=100~\mu s-700~\mu s$ in Eq.~\ref{eq23} for a fixed $\rho(\textbf{M}) = 0$ is also plotted in blue in Fig.~\ref{fig3}. It is evident that the spatial model mismatch has  detrimental effect on the achievable bandwidth of the COFB system for the disturbance rejection. The nature of this bandwidth reduction is different in comparison to the pure delay. There is a weak reduction in bandwidth when $\rho(\textbf{M})$ moves from 0 to 0.2, and the effect is much sharper when $\rho(\textbf{M})$ approaches unity. These results are very helpful to decide the required ORM update rate in the COFB controller for a variable optics machine e.g. for the on-ramp orbit correction in booster rings. For a given dynamical system and for the required bandwidth, one can calculate the allowed range of $\rho(\mathbf{M})$ using Eq.~\ref{eq23} which can be kept below a certain threshold by updating the $\mathbf{R}_{\Theta}^{+}$ in the controller after appropriate time steps. 
\\
\subsection{Closed-loop stability}
\begin{figure}[ht]
    \centering
    \includegraphics*[width=8.3cm]{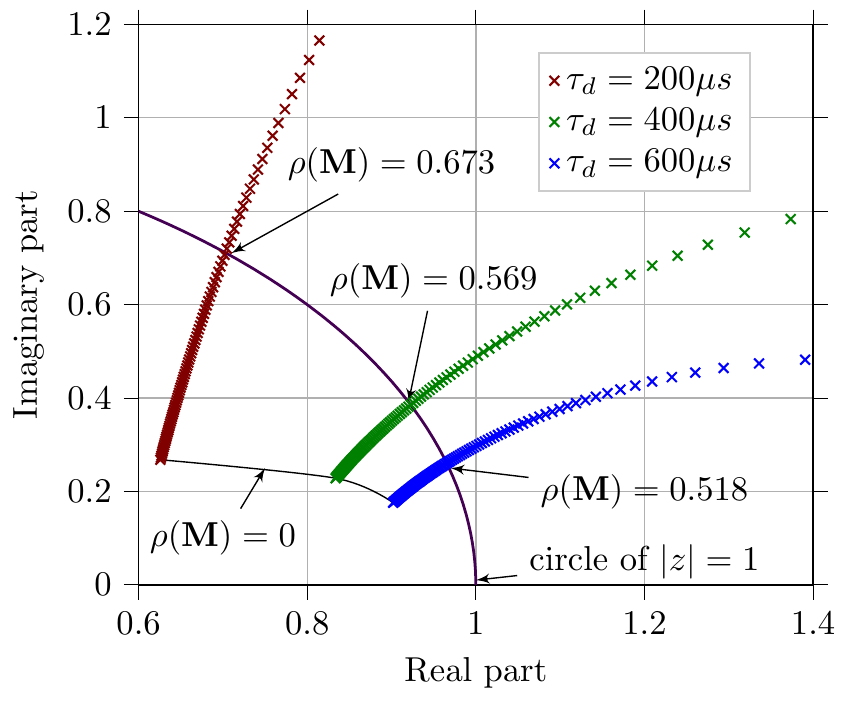}
    \caption{The pole movement over the complex plane from inside to outside of the unit circle with the increase of $\rho(\textbf{M})$}.
    \label{fig14}
\end{figure}

Equation~\ref{eq25} suggests that for the fast regime of the orbit correction, $\rho(\textbf{M}) > 1$ will change the sign of the controller gains resulting in the closed-loop instability. This is regardless of the dynamics of the system and the parameters of the controller and is consistent to the instability condition of slow regime given in inequality~\ref{eq7}. \\
Eq.~\ref{eq27} provides precious information about the required controller parameter variation for a COFB system where the machine optics changes over time but the ORM is not updated accordingly i.e. $\rho(\mathbf{M})$ increases over time from $0$ to $1$. In this case, the controller parameters are needed to be scaled with ($1-\rho(\textbf{M})$) relative to the controller parameters corresponding to $\rho(\textbf{M}) = 0$. \\ 
In order to demonstrate the need of such a scaling, three systems are taken with bandwidth of $850$~Hz and delays $\tau_d= 200,400$ and $600~\mu s$. The PI controllers for $\rho(\mathbf{M})=0$ are designed using IMC-tuning~\cite{IMC_PI_course} and the poles of the complementary sensitivity function $T_{\rho}(z)=\frac{L_\rho(z)}{1+L_\rho(z)}$ are calculated as a function of $\rho(\mathbf{M})$ while keeping the controller parameters fixed corresponding to $\rho(\mathbf{M})=0$. Fig.~\ref{fig14} shows the dominant pole positions of $T_{\rho}(z)$ on the complex plane for all three systems. The poles move from inside to the outside (unstable region) of unit circle defined for $|z|=1$ with an increase of $\rho(\textbf{M})$ for all three cases whereas the cross-over value of $\rho(\textbf{M})$ depends upon the dynamics (delays in this case) of the system.\\
These simulations show that the COFB can become unstable if the magnitude of the model mismatch is not properly investigated along with the system dynamics and should be taken into account of the controller parameters in order to run the COFB system in the fixed-ORM mode when the machine optics varies over time. The use of the controller gains corresponding to the $\rho(\textbf{M}) = 0$ can lead to the system instability at a characteristic frequency defined by the dynamics of the system. The conclusions drawn from this study have been helpful in the first beam-based test of the fast COFB system of SIS18 in the $x$-plane~\cite{ibic19}. 
It is important to mention that the instability predicted for $\rho(\textbf{M})<1$ in Fig.~\ref{fig14} is not in contradiction to the condition of inequality~\ref{eq7} which is defined for the slow regime where the dynamics of the system can be ignored and there is no need of scaling of the controller parameters. One can see that the threshold of $\rho(\mathbf{M})$ in Fig.~\ref{fig14} moves towards $1$ as the delay of the system decreases relative to the sampling time $T_s=100~\mu s$. On the other hand, if the controller parameters are scaled according to Eq.~\ref{eq27} as a function of model mismatch in the fast regime, the COFB instability due to the model mismatch will occur only for $\rho(\mathbf{M})>1$, resulting in the identical condition for instability for both slow and fast regimes.   

\section{Summary and conclusions}
In this paper, the effect of spatial model mismatch on the stability and the performance of the COFB systems is investigated for the slow and the fast regimes of orbit correction, separately. In the slow regime, the temporal dynamics of the system can be ignored in comparison to the low rate of orbit correction and the closed-loop response depends only on the orbit response matrices of the controller and the system. For this regime, the spectral radius condition of stability is defined for the correction matrix $\textbf{M}$ which yields a higher and practical stability margin in comparison to the previously used small gain theorem, against the spatial model mismatch. It is shown that the stability margins yielded by the two conditions can be related through the condition number of the nominal (controller) ORM, with exact equality for the circulant symmetry. A relation of the spectral radius $\rho(\textbf{M})$ with the correction ratio is also discussed.
The spectral radius is calculated for various sources of model mismatch in SIS18 in order to obtain hints for the optimal design of a COFB system. For on-ramp ORM variation due to triplet to doublet quadrupole transition, it is shown with the help of simulations in MAD-X that the defined spectral radius is a practical figure of merit for quantifying the spatial model mismatch. The simulations predicted that SIS18 COFB system will not be unstable even for the maximum on-ramp model mismatch for the slow regime of operation in both planes. The calculations of spectral radius are also performed for intensity-dependent tune shift and the beta-beating in SIS18. It is shown that the spectral radius is related to the magnitude of the tune shift and more sensitive in the direction towards the integer value of tune, and hence the instability threshold to spatial model mismatch depends on the working point. Further, it is found that beta beating itself is not a crucial source of model mismatch unless accompanied by a tune shift towards an integer value. The conclusions drawn from the calculations and simulations for SIS18 were experimentally verified at COSY FZJ where the closed orbit is made unstable with artificially induced tune shift-based model mismatch. The observed instability threshold is in agreement with the instability threshold predicted by the spectral radius condition. \\
These result from the slow regime are extended to the fast feedback regime also accounting for closed-loop dynamics. It is found that the spectral radius condition defines the stability condition of the fast COFB system for a given temporal system model. The controller settings must be scaled with $\rho(\textbf{M})$ for a fixed-ORM mode of the COFB system for a variable optics machine, in order to ensure the closed-loop stability. The achievable system bandwidth can be calculated from $\rho(\textbf{M})$ and in order to achieve a required bandwidth, $\rho(\textbf{M})$ can be controlled by the appropriate updating of the ORM in the contorller of the COFB system.
In nutshell, a single measurable parameter, i.e. spectral radius of the correction matrix $\textbf{M}$ quantifying the model mismatch is enough for predicting the residual orbit, bandwidth reduction and the stability criteria in terms of controller parameters for the feedback system in consistency with the practical limits.


\begin{thebibliography}{10}
\expandafter\ifx\csname url\endcsname\relax
  \def\url#1{\texttt{#1}}\fi
\expandafter\ifx\csname urlprefix\endcsname\relax\def\urlprefix{URL }\fi
\expandafter\ifx\csname href\endcsname\relax
  \def\href#1#2{#2} \def\path#1{#1}\fi
  
 \bibitem{gen}
G.~Rehm, {Achieving and measuring sub-micrometer beam stability at $3^{rd}$ generation light sources}, J. Phys. Conf. Ser. $\bf{425}$, 1 (2013).
\bibitem{cofbreview}
D.~Bulfone, {Overview of fast beam positon feedback systems}, CERN report 1183171, (2008).
\bibitem{sandira1}
S.~Gayadeen, {Synchrotron Electron Beam Control}, Ph.D. thesis, Department of Engineering Science, University of Oxford, UK, 2014.
\bibitem{nsls}
W.~Cheng, {Chracterization of NSLS2 storage ring beam orbit stability}, in \textit{Proceedings of the 4$^{th}$ International Beam Instrumentation Conference (IBIC 2015), Melbourne, Australia, 2015}, p. 625. 
\bibitem{esrf}
E.~Plouviez, F.~Epaud, J.~M.~Koch, K.~B.~Scheidt {The new fast orbit correction system of the ESRF storage ring}, in \textit{Proceedings of the $10^{th}$ European Workshop on Beam Diagnostics and Instrumentation for Particle Accelerators (DIPAC 2011), Hamburg, Germany, 2011}, p. 215. 
\bibitem{petra}
A.~Brenger, I.~Krouptchenkov, G.~Kube, F.~Schmidt-F\"ohre and K.~Wittenburg, {Experience with the commissioning of the Libera brilliance BPM electronics at PETRA III}, in \textit{Proceedings of the $9^{th}$ European Workshop on Beam Diagnostics and Instrumentation for Particle Accelerators (DIPAC 2009), Basel, Switzerland, 2009}, p. 215.
\bibitem{ducane}
S. R. Duncan {The design of a fast orbit beam stabilisation system for the Diamond
synchrotron}, Technical Report 2296/07, Department of Engineering Science,
University of Oxford, 2007.
\bibitem{sami}
M. S. Fadali and A. Visioli, \textit{Digital Control Engineering} (Academic Press, Boston, 2009), Chap. 2, p. 55-86.
\urlprefix\url{http://www.sciencedirect.com/science/article/pii/B9780123744982000035}.


\bibitem{Sands}
M.~Sands, {The physics of electron storage rings: An introduction}, Stanford Linear Accelerator Center Report No. SLAC- 121, 1970.
\bibitem{svd1}
Y.~Chung, G.~Decker, and K.~Evans, Closed orbit correction using singular value decomposition of the response matrix, Argonne National Laboratory, IL, 1993 (unpublished).
\bibitem{svdgolub}
G.~H.~Golub and C.~F.~V.~Loan, \textit{Matrix Computations} (The Johns Hopkins University Press, 2013), Chap. 2.

\bibitem{sandira2}
S.~Gayadeen, {Uncertainty modelling of response matrix}, in \textit{Proceedings of the $15^{th}$ International Conference on Accelerators and Large Experimental Physics Control Systems (ICALEPCS 2015), Melbourne, Australia, 2015}, p. 506.
\bibitem{self1}
S.~H.~Mirza, P.~Forck, H.~Klingbeil and R.~Singh, {Investigations of spatial
  process model for the closed orbit feedback system at the SIS18 synchrotron
  at GSI}, in \textit{Proceedings of the $16^{th}$ International Conference on Accelerators and Large Experimental Physics Control Systems (ICALEPCS 2017), Barcelona, Spain, 2017}, p. 1301.
\bibitem{self2}  
S.~H.~Mirza, R.~Singh, P.~Forck, H.~Klingbeil {Closed orbit correction at synchrotrons for symmetric and near-symmetric lattices}, Phys. Rev. ST Accel. Beams, $\bf{22}$, 072804 (2019).
\bibitem{self3}
S.~H.~Mirza, P.~Forck, H.~Klingbeil and R.~Singh, {Effect of model errors on the
  closed orbit correction at the SIS18 synchrotron of GSI}, J. Phys. Conf. Ser. $\bf{1067}$ 2080 (2018).

\bibitem{bonn}
M.~Eberhardt, {Measurement and correction of the longitudinal and transverse tune during the fast energy ramp at ELSA}, in \textit{Proceeding of the $1^{st}$ International Particle Accelerator Conference (IPAC 2010), Kyoto, Japan, 2015}, p. 897.
\bibitem{RS2}
R.~Singh, O.~Boine-Frankenheim, O.~Chorniy, P.~Forck, R.~Haseitl, W.~Kaufmann, P.~Kowina, K.~Lang, and T.~Weiland, {Interpretation of transverse tune spectra in a heavy-ion synchrotron at high intensities}, Phys. Rev. ST Accel. Beams, 16, 034201, (2013).
\bibitem{tdb}
O.~ Kester, P.~Spiller and H.~Stoecker, {FAIR project at GSI in Challenges and Goals for Accelerators in the XXI Century}, (2016), 611-622, WORLD SCIENTIFIC.
\bibitem{sgt}
M.~Green and D. J. N. Limebeer. {Linear Robust Control}, (Prentice Hall,1994), Ch. 3,  ISBN:
0-131-02278-4.
\bibitem{cernuncern}
P.~J\"urgen and S.~Jochem, {Analytic robust stability analysis of SVD orbit feedback}, CERN-OPEN-2012-019 (2012).
\bibitem{multivariable}
S.~Skogestad and I.~Postlethwaite, {Multivariable feedback control; analysis and design}, (John Wiley and Sons, 2004), Ch. 8. 
\bibitem{accelerators}
O.~Bruning and S.~Myers \textit{Challenges and Goals For Accelerators In The XXI Century},
(World Scientific, 2016, 287-289).
\bibitem{cosy}
R.~Maier, S.~Martin and U.~Pfister, {The COSY-J\"ulich project-Feb. 1989 status} in \textit{Proceedings of the Particle Accelerator Conference, 1989} vol.1. 40 - 42. 
\bibitem{Eigvalue}
B.~N.~Parlett, \textit{The Symmetric Eigenvalue Problem}, (University of California, Berkeley, California, 1998).
\bibitem{lax}
P.~D.~Lax, {Functional Analysis}, (Wiley-Interscience, 2002), p.195. ISBN 0-471-55604-1
\bibitem{FAIR}
FAIR - Baseline Technical Report, GSI, 2006.
\bibitem{Spiller}
P. Spiller, {Statuts of the fair synchrotron projects SIS18 upgrade and SIS100}, in {Proceedings of the $5^{th}$ International Particle Accelerator Conference (IPAC 2014), Dresden,
Germany, 2014}, p. 1857.
\bibitem{RS1}
R. Singh, {Tune measurement at GSI SIS-18: Methods and Applications}, PhD Thesis, TU Darmstadt (2014).
\bibitem{madx}
H.~Grote, F.~Schmidt, L.~Deniau and G.~Roy, Computer code MAD-X: user’s reference manual, European  Organization  for  Nuclear  Research, 2016.
\bibitem{ZOTTER}
B.~Zotter, {The Q-shift of off-center particle beams in elliptic vacuum chambers}", Nucl. Instrum. Methods Phys. Res. Sect. A \textbf{129}, 377-395, (1975). 
\bibitem{bb}
K.~Wille, {The Physics of Particle Accelerators, an introduction} (Oxford University Press, 2000), Chap. 3.15.
\bibitem{IMC}
B. Francis and W. Wonham {The Internal Model Principle of Control Theory} Automatica. 12. 457-465(1976). \bibitem{michael}
M.~Abbott, {Using an Internal model controller for electron beam position fast feedback }, Diamond Light Source, internal document, (2007).
\bibitem{IMC_PI_course}
{THE IMC-BASED PID PROCEDURE}
https://rpi.edu/dept/chem-eng/WWW/faculty/bequette/courses/cpc/IMC\_PID.pdf. 
\bibitem{IMC_PI}
I-Lung~Chien, {
IMC-PID Controller Design - An Extension}, IFAC Proceedings Volumes, Volume 21, Issue 7, 1988,
Pages 147-152,
ISSN 1474-6670,
https://doi.org/10.1016/S1474-6670(17)53816-1.
\bibitem{ibic19}
R.~Singh et al., {First beam-based test of fast closed orbit feedback system at GSI SIS18}in {Proceedings of the $6^{th}$ International Beam Instrumentation Conference (IBIC 2019), Malmoe,
Sweden, 2019}, p. 1857.

\end{thebibliography}
\end{document}